\documentclass[twocolumn,english,aps,superscriptaddress,prl,floatfix,nolongbibliography,footinbib]{revtex4-2}
\usepackage[latin9]{inputenc}
\usepackage{color}
\usepackage{babel}
\usepackage{amsmath}
\usepackage{amssymb}
\usepackage{graphicx}
\usepackage{bm}
\usepackage{dsfont}
\usepackage{physics}
\usepackage{tikz}
\usepackage[compat=1.1.0]{tikz-feynman}
\usepackage[unicode=true,pdfusetitle,bookmarks=true,bookmarksnumbered=false,bookmarksopen=false,breaklinks=true,pdfborder={0 0 0},pdfborderstyle={},backref=false,colorlinks=true]{hyperref}
\hypersetup{linkcolor=blue,citecolor=blue,urlcolor=blue}

\newcommand{\hc}{\mathrm{H.c.}}

\newcommand{\chit}{\tilde{\chi}}

\newcommand{\ii}{\mathrm{i}}

\newcommand{\HH}{\mathcal{H}}

\renewcommand{\vec}{\mathbf}
\newcommand{\eps}{\varepsilon}

\graphicspath{{./}{./figs/}}

\begin{document}

\title{Heavy-fermion hybridization waves from Dirac spinons\\near a Kondo-breakdown quantum phase transition}

\author{Pedro M. C\^onsoli}
\affiliation{Department of Physics, Arizona State University, Tempe, AZ 85287, USA}
\author{Matthias Vojta}
\affiliation{Institut f\"ur Theoretische Physik and W\"urzburg-Dresden Cluster of Excellence ctd.qmat, Technische Universit\"at Dresden, 01062 Dresden, Germany}
\author{Onur Erten}
\affiliation{Department of Physics, Arizona State University, Tempe, AZ 85287, USA}


\begin{abstract}
	Motivated by the recent observation of spatially modulated screening in different paramagnetic heavy-fermion materials, we propose a theoretical mechanism for the emergence of such hybridization waves in frustrated Kondo lattices with no magnetic order. By applying both a parton mean-field theory and a perturbative calculation to a minimal Kondo-Heisenberg lattice model, we demonstrate that a heavy Fermi liquid can naturally develop hybridization waves when it approaches a quantum phase transition where Kondo screening breaks down and the local magnetic moments form a Dirac spin liquid. We relate the strength of this mechanism to a kinematic condition on the electronic and spinon band structures and show that the spatial periodicity of the ensuing hybridization wave is controlled by momenta $\vec{Q}$ that connect the Dirac nodes in the spinon dispersion to low-energy portions of the electronic band structure.
\end{abstract}

\date{\today}
\maketitle


\paragraph{Introduction---} Heavy-fermion materials have long served as a fruitful platform for the study of complex many-body phenomena, including unconventional forms of superconductivity, quantum criticality, and non-Fermi liquid behavior \cite{stewart01,loehneysen07,paschen21}.
Typically, the physics of such systems is rooted in the interplay between conduction electrons and local magnetic moments formed by strongly correlated $f$ electrons. The characteristic enhancement of the quasiparticle mass, in particular, originates from the coherent screening of the $f$ moments by the conduction electrons, known as the lattice Kondo effect \cite{hewson_book}.

\begin{figure}[t]
	\centering
	\includegraphics[width=0.95\columnwidth]{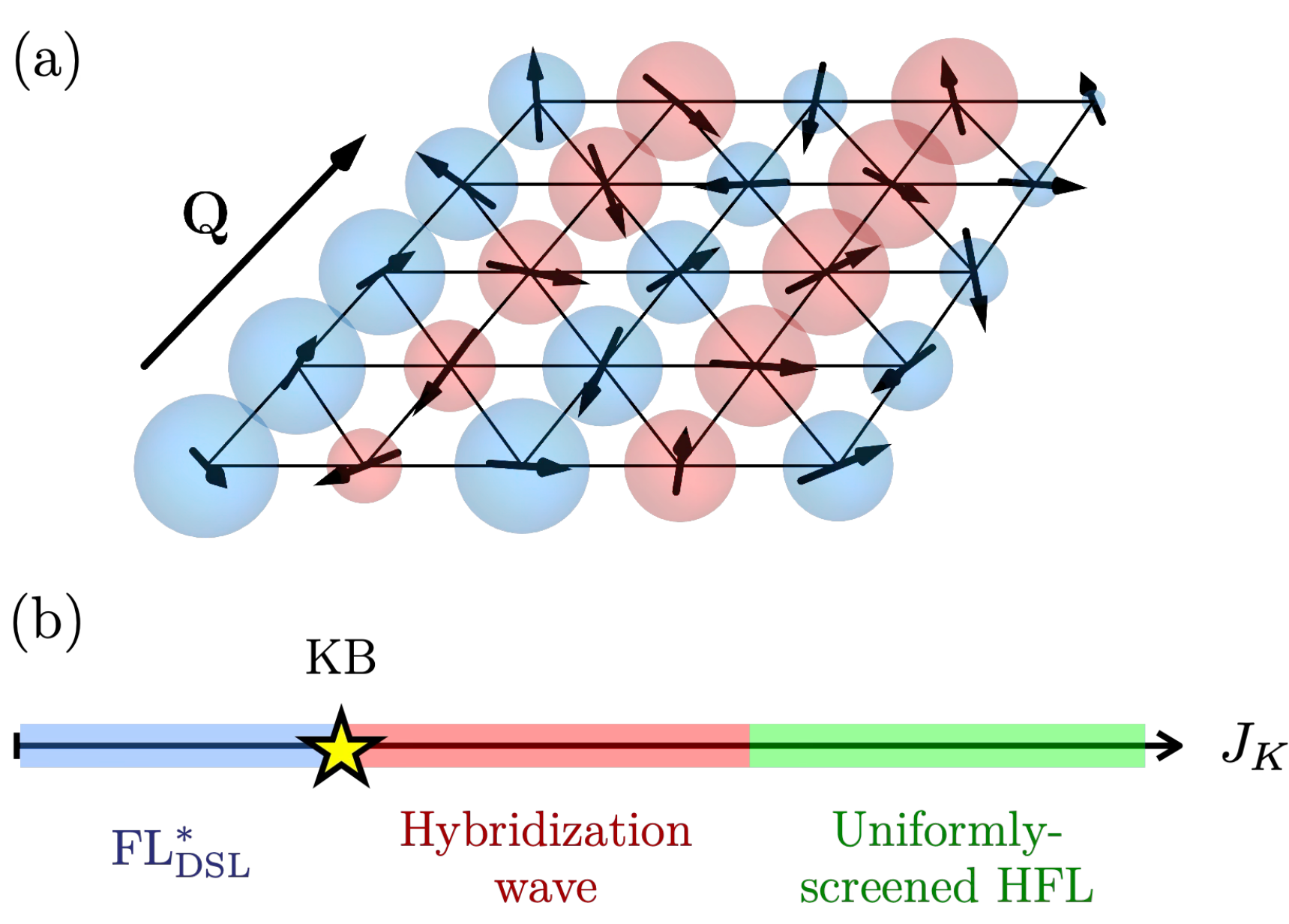}
	\caption{(a) Kondo hybridization wave (HW) with a two-sublattice unit cell and ordering wavevector $\vec{Q}$ on the triangular lattice. The spatial profile of Kondo screening is encoded in the radii of the spheres surrounding the local magnetic moments. Blue and red spheres mark the different sublattices. (b) Schematic zero-temperature phase diagram of a Kondo lattice as a function of the Kondo coupling $J_K$. The HW emerges at intermediate couplings, between a fractionalized Fermi liquid with a Dirac spin-liquid component (FL$^*_\mathrm{DSL}$) and a uniformly-screened heavy Fermi liquid. The yellow star marks the Kondo-breakdown (KB) transition.}
	\label{fig:schem}
\end{figure}

While the onset of Kondo screening with decreasing temperature is usually a crossover, it can also constitute a genuine phase transition if accompanied by the spontaneous breaking of a physical symmetry. The latter applies, for instance, to hastatic \cite{chandra13,chandra15} and nematic \cite{vijayvargia24, jiang26} heavy-fermion states, which feature broken time-reversal and lattice rotational symmetries, respectively.
Kondo hybridization waves (HWs) represent another example, in which Kondo screening develops periodic modulations that break lattice translational and rotational symmetries [see Fig.~\ref{fig:schem}(a)]. Early studies \cite{lacroix96,regueiro97,motome10} proposed that a version of this phenomenon, termed partial Kondo screening, can serve as a mechanism to relieve frustration in metallic heavy-fermion antiferromagnets such as UNi$_4$B \cite{mentink94,oyamada07} and CePdAl \cite{doeni96,oyamada08}. Subsequent work suggested that HWs can also play a role in \emph{paramagnetic} heavy-fermion materials. This was first put forward (but not confirmed) as an explanation for the hidden order in URu$_2$Si$_2$ \cite{dubi11,su11,herrera21}; more recently, paramagnetic HWs were detected in scanning tunneling microscopy experiments on UTe$_2$ \cite{yu26} and 6R-TaS$_2$ \cite{cao26}.
However, contrary to partial Kondo screening, the physical mechanism underlying HWs in the paramagnetic setting remains uncertain. To our knowledge, the only concrete scenario advanced thus far is that HWs may arise when a heavy Fermi liquid approaches a quantum phase transition to a fractionalized Fermi liquid (FL$^*$) \cite{senthil03,senthil04,senthil05} where the $f$ moments decouple from the conduction electrons to form a U(1) quantum spin liquid with a spinon Fermi surface \cite{paul07,zhu08}. In this picture, the periodicity of a HW is roughly set by momentum-space separation between the spinon and electronic Fermi surfaces. However, spinon-Fermi-surface states appear rare, both experimentally and theoretically.

In this Letter, we establish an alternative scenario, Fig.~\ref{fig:schem}, whereby HWs emerge on the brink of a Kondo-breakdown (KB) transition to an FL$^*$ phase featuring a Dirac spin liquid (DSL), as opposed to a spinon-Fermi-surface state, with a concrete realization in a triangular-lattice setting. Our findings acquire particular significance in view of the advent of high-tunable moir\'e Kondo lattices \cite{vano21,kumar22,zhao23} and the growing theoretical evidence that frustrated triangular-lattice Heisenberg models realize a DSL \cite{iqbal16,hu19,sherman23,wietek24}.


\paragraph{Kondo-Heisenberg lattice model---} We consider a minimal Kondo-Heisenberg model on a triangular lattice with Hamiltonian $\HH = \HH_t + \HH_K + \HH_H$ and
\begin{subequations}
	\label{eq:khlm}
	\begin{align}
		\HH_t &= -\sum_{i,j} \sum_{\alpha = \uparrow, \downarrow} t_{ij} c_{i\alpha}^\dagger c_{j\alpha},
		\label{eq:Ht}
		\\
		\HH_K &= J_K \sum_{i} \sum_{\alpha, \beta} \vec{S}_{i} \cdot \left( c_{i\alpha}^\dagger \frac{\bm{\sigma}_{\alpha\beta}}{2} c_{i\beta} \right),
		\label{eq:HK}
		\\
		\HH_H &= J_H \sum_{\langle ij \rangle} \left[
		\vec{S}_{i} \cdot \vec{S}_{j} + \frac{\kappa}{2} \left(\vec{S}_{i} \cdot \vec{S}_{j}\right)^2
		\right].
		\label{eq:HH}
	\end{align}
\end{subequations}
Equations~\eqref{eq:Ht} and \eqref{eq:HK} account for the hopping of the conduction electrons and their Kondo coupling to spin-$1/2$ moments $\vec{S}_i$, while Eq.~\eqref{eq:HH} contains two types of interactions between nearest-neighbor moments: an antiferromagnetic Heisenberg exchange and a biquadratic coupling whose strength is set by $\kappa$.
Although the biquadratic term reduces to bilinear Heisenberg interactions for spin-$1/2$ SU$(2)$ moments, and thus only renormalizes $J_H$, it plays a more significant role when the spin symmetry is enlarged to SU$(N)$---an ingredient we shall use in the SU$(N)$-based mean-field theory below.
For concreteness, we restrict ourselves to nearest-neighbor $c$-electron hopping with uniform amplitude $t$, which we set as our unit of energy. Our results can be generalized to capture more realistic electronic dispersions.

We shall focus on the low-temperature physics of Eq.~\eqref{eq:khlm}, which is governed by the competition between two energy scales: the Kondo temperature $T_K$ and the intermoment exchange $J_H$~\cite{doniach77}. When $T_K \gg J_H$, the coherent screening of local moments via the lattice Kondo effect results in a Kondo insulator or a heavy Fermi liquid. Conversely, when $T_K \ll J_H$, Kondo screening is suppressed and the local moments either order magnetically or form a quantum spin liquid. The latter scenario gives rise to an FL$^*$ phase, whose existence in different microscopic models has been demonstrated via nonperturbative methods~\cite{tsvelik16,hofmann19,danu20,wang20}.


\paragraph{Parton mean-field theory---} We investigate Eq.~\eqref{eq:khlm} by employing the Abrikosov fermion representation, $\vec{S}_i = \frac{1}{2} \sum_{\alpha,\beta} f_{i\alpha}^\dagger \bm{\sigma}_{\alpha\beta} f_{i\beta}$, subject to the constraint $\sum_{\alpha} f_{i\alpha}^\dagger f_{i\alpha} = 1$, which is enforced by a dynamical field $\lambda_i$ in the framework of a coherent-state path integral \cite{read83,coleman_book}.
We then decouple $\HH_K$ and $\HH_H$ via Hubbard-Stratonovich transformations that introduce a pair of complex bosonic fields, $b_i$ and $\chi_{ji}$, conjugate to $c_{i\sigma}^\dagger f_{i\sigma}$ and $f_{i\sigma}^\dagger f_{j\sigma}$ \cite{coleman83,read84,affleck88,marston89,suppl}. The saddle-point equations of the resulting action define a mean-field theory that becomes exact when the spin symmetry is enhanced to SU$(N)$ and $N\to \infty$.

\begin{figure*}
	\centering
	\includegraphics[width=\textwidth]{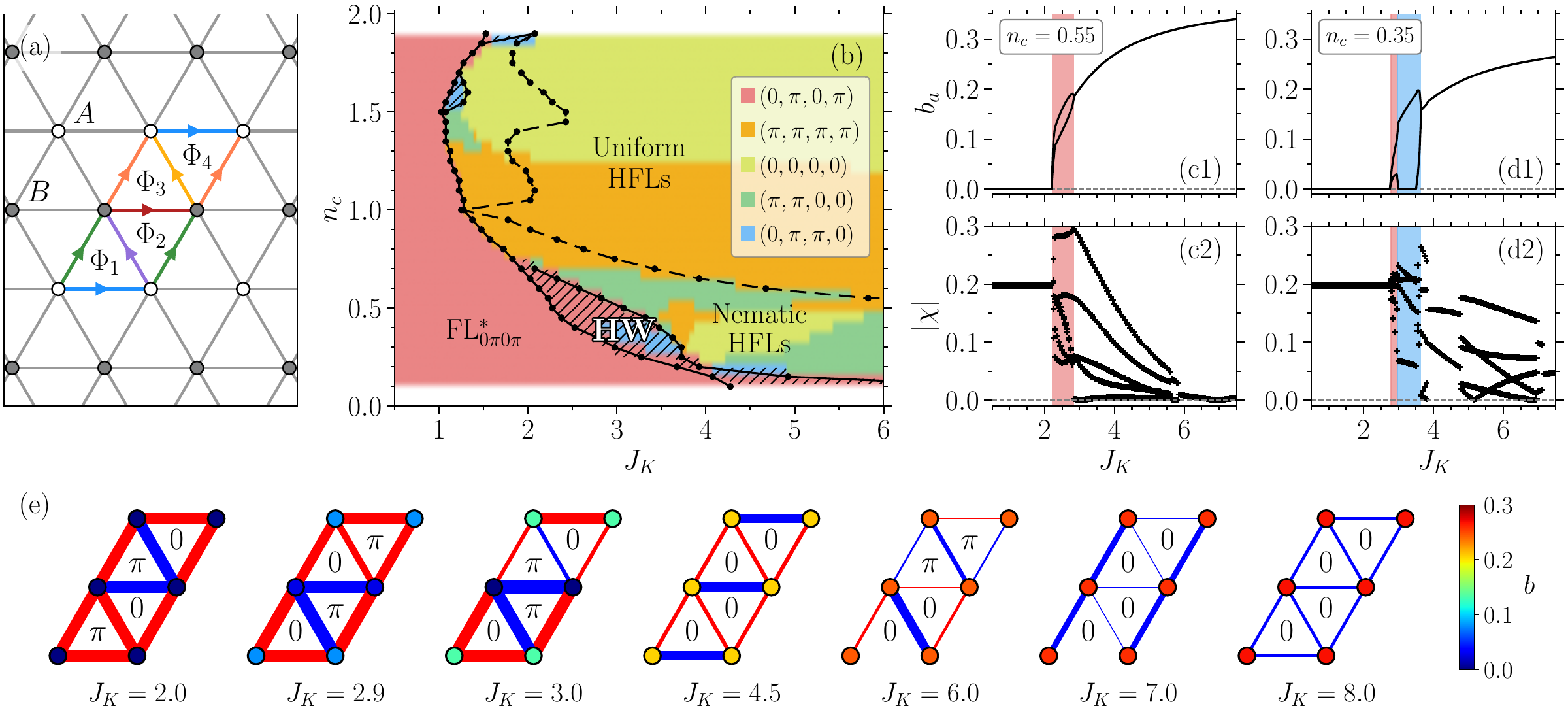}
	\caption{(a) Triangular lattice with sublattices $A$ and $B$. The unit cell of our mean-field calculations is illustrated in the center, where we have labeled the four independent fluxes, $\Phi_1,\ldots, \Phi_4$, and represented the six $\chi$ fields as directed bonds of different colors. (b) Mean-field phase diagram for $L=150$, $J_H = 0.2$, $\kappa=1$, and $k_\mathrm{B} T = 10^{-3}$. The hatched portions correspond to hybridization waves (HW), which arise in the proximity of a fractionalized Fermi liquid (FL$^*_{0\pi0\pi}$). The dashed lines separate regions with spatially uniform (right) and nematic (left) absolute values of the $\chi$ fields. Different flux sectors are distinguished by the colors in the background. (c,d) Numerical results for the $b_a$ and $\chi$ fields along two horizontal cuts, (c) $n_c = 0.55$ and (d) $0.35$. Intervals supporting HWs are highlighted in color, according to the flux sector of the state. (e) Visualization of mean-field solutions as a function of $J_K$ for $n_c = 0.35$. The absolute value and sign of each $\chi$ variable are encoded in the thickness and color (blue for positive, red for negative) of the corresponding bond. Sites are also colored according to the magnitude of the local $b$ field.}
	\label{fig:mfplots}
\end{figure*}

Here, we study a version of the above mean-field theory that assumes invariance under translations of a \emph{two-site} unit cell [see Fig.~\ref{fig:mfplots}(a)], chosen to match the real-space modulations observed in the STM experiments of Ref.~\onlinecite{cao26}. Larger mean-field unit cells are also possible, but will not be considered here for simplicity. A treatment of HWs with larger spatial periods will thus be deferred to the perturbative calculations presented at the end of the paper.
Labeling each site by its unit cell $i$ and sublattice $a = A,B$, we can write the mean-field Hamiltonian as
\begin{align}
	\HH_\mathrm{MF}
	&= E_0 + \HH_t + \sum_{ia\sigma} \lambda_a f_{ia\sigma}^\dagger f_{ia\sigma}
	\notag \\
	&- J_K \sum_{ia\sigma} \left( b_a c_{ia\sigma}^\dagger f_{ia\sigma} + \hc \right)
	\notag \\
	&- J_H \sum_{\langle ia, jb \rangle, \sigma} \left( \tilde{\chi}_{jb, ia} f_{ia\sigma}^\dagger f_{jb\sigma} + \hc
	\right),
	\label{eq:hmf1}
\end{align}
where $E_0$ is a constant, $\lambda_a$ are static Lagrange multipliers, and $\tilde{\chi}_{jb, ia} = ( 1 - 2\kappa \abs{\chi_{jb, ia}}^2 ) \chi_{jb, ia}$. While complex in principle, the mean fields $b_a = \expval{ f_{ia\sigma}^\dagger c_{ia\sigma} }$ can be rendered real and positive via sublattice-dependent gauge transformations. Their absolute values then encode the strength of Kondo screening on each sublattice.
The six independent bond variables $\chi_{jb, ia} = \expval{ f_{jb\sigma}^\dagger f_{ia\sigma} }$ give rise to nine gauge-invariant quantities: six absolute values and three independent fluxes among $\Phi_{1}, \ldots, \Phi_4$ \cite{flux_note}.  The former quantify the delocalization of $f$ spinons along nearest-neighbor bonds, whereas the latter correspond to the complex phases the spinons accumulate after circulating clockwise around the elementary triangular plaquettes in Fig.~\ref{fig:mfplots}(a).
Further details on the derivation of Eq.~\eqref{eq:hmf1} and the accompanying self-consistency equations are provided in the Supplemental Material (SM) \cite{suppl}.

\paragraph{Pure local-moment sector---} Various studies of antiferromagnetic Heisenberg models with $\kappa=0$ have demonstrated that the large-$N$ mean-field theory has a tendency toward dimerized valence-bond-solid (VBS) order~\cite{affleck88,read89a,read89b,assaad05,schwab23}. By solving the self-consistency equations in the pure $f$ sector \cite{suppl}, we confirm that our triangular-lattice model reproduces this trend.
Since the resulting VBS ground state is unrealistic in the physical SU(2)-symmetric case, we follow Refs.~\onlinecite{marston89} and \onlinecite{drechsler25} by introducing a biquadratic interaction as a practical knob to suppress dimerization and stabilize a competing spin liquid instead. As shown in the SM \cite{suppl}, this goal is achieved by a moderate $\kappa \gtrsim 0.3$, which gives rise to a gapped chiral spin liquid (CSL) with uniform $\Phi_n = \pi/2$ fluxes.

Numerical work on a related triangular-lattice model---the antiferromagnetic $J_1$-$J_2$ SU(2) Heisenberg model---has put forward strong evidence for the existence of a quantum-disordered ground state at intermediate $J_2/J_1$ which appears, however, to be a U(1) DSL instead of a CSL~\cite{iqbal16,hu19,sherman23,wietek24}. Hence, we sought an adjustment of the mean-field theory capable of reproducing a DSL ground state. After verifying that the addition of $J_2$ interactions was insufficient, we opted to restrict the theory to \emph{real} $\chi$ variables, thereby eliminating the CSL and all other time-reversal-symmetry-breaking solutions. As a result, the system transitions from the VBS phase to a DSL with $(\Phi_1, \Phi_2, \Phi_3, \Phi_4) = (0,\pi,0,\pi)$ at $\kappa \approx 0.7$ \cite{suppl}; this is precisely the flux structure found in numerical studies of the $J_1$-$J_2$ model \cite{iqbal16}.
%

\paragraph{Mean-field phase diagram---} To analyze the predictions of our mean-field theory, we construct phase diagrams at fixed $J_H$ and $\kappa$ as a function of $J_K$ and the filling $n_c = \frac{1}{N_s} \sum_{ia\sigma} \expval{c_{ia\sigma}^\dagger c_{ia\sigma}}$, where $N_s$ denotes the total number of sites.
We consider systems with $L \times (L/2)$ unit cells along the primitive vectors, $\vec{a}_1 = (1,0)$ and $\vec{a}_2 = (1,\sqrt{3})$, subject to periodic boundary conditions. For consistency, we limit ourselves to solutions described by real $b$ and $\chi$, thereby extending the restriction to time-reversal-symmetric states to the entire phase diagram. We will nonetheless abandon this constraint in the perturbative calculations presented at the end of the paper.

Figure~\ref{fig:mfplots}(b) shows the phase diagram for $L=150$, $J_H = 0.2$, $\kappa=1$, and $k_\mathrm{B} T=10^{-3}$. Results for other values of $J_H$ and $\kappa$ are presented in the SM \cite{suppl}.
In the weak-coupling limit, $J_K \to 0$, the system realizes an FL$^*_{0\pi0\pi}$ phase in which light electron quasiparticles coexist with the DSL described above. The absence of Kondo screening manifests in the condition $b_A = b_B = 0$, which implies that the local moments do not contribute to the volume enclosed by the electronic Fermi surface \cite{oshikawa00,senthil04}.
In the opposite limit,  $J_K \to \infty$, we verify a tendency toward homogeneous heavy Fermi liquids, with $b_A = b_B \ne 0$ and $\abs{\chi_{jb, ia}} = \chi > 0$ [see Figs.~\ref{fig:mfplots}(c-e)]. The optimal flux configuration at a given $n_c$ corresponds to the set of unhybridized $f$ bands that yields the most favorable hybridized band structure. This is found to be $(\Phi_1, \Phi_2, \Phi_3, \Phi_4) = (\pi,\pi,\pi,\pi)$ around half filling, but to change to $(0,0,0,0)$ as $n_c \to 2$.

Upon decreasing $J_K$ at fixed $n_c \ne 1$, we observe a transition into a nematic regime where nonuniform bond variables $\abs{\chi_{jb, ia}}$ break the rotational symmetry of the triangular lattice, but the hybridization fields remain uniform and nonzero, $b_A = b_B \ne 0$.
A further decrease of $J_K$ leads to one of two outcomes, depending on $n_c$. The first is a direct first-order phase transition into the FL$^*_{0\pi0\pi}$ phase, marked by an abrupt drop of $b_A$ and $b_B$ to zero and a change in the flux sector. Alternatively, upon entering either of the hatched regions in Fig.~\ref{fig:mfplots}(b), a phase with inhomogeneous Kondo screening ($b_A \ne b_B$) and nematic $\abs{\chi_{jb, ia}}$ emerges.
Within our mean-field theory, there are two variants of such HWs, denoted $\mathrm{HW}_{0\pi\pi0}$ and $\mathrm{HW}_{0\pi0\pi}$ to reflect that their flux configurations are $(0,\pi,\pi,0)$ and $(0,\pi,0,\pi)$, respectively.
Interestingly, $\mathrm{HW}_{0\pi\pi0}$ states with $n_c < 1$ always have one vanishing $b$ field, so that the corresponding sublattice is \emph{completely} unscreened. Furthermore, because this phase has a different flux configuration than the FL$^*_{0\pi0\pi}$, it cannot undergo a \emph{continuous} KB transition.
However, no such restriction applies to the $\mathrm{HW}_{0\pi0\pi}$ phase, for which the $b$ fields vanish continuously on approaching the boundary to the FL$^*_{0\pi0\pi}$ [see Fig.~\ref{fig:mfplots}(c)].

Our analysis raises the question of how the emergence of HWs correlates with the filling. A first hint to the answer comes from the evolution of the bare $c$-electron Fermi surface with $n_c$. As shown in the SM \cite{suppl}, the Fermi surfaces cross the nodes in the dispersion of the DSL for $n_c \approx 0.45$, which lies at the center of a large region in Fig.~\ref{fig:mfplots}(b) with a continuous FL$^*_{0\pi0\pi}$--$\mathrm{HW}_{0\pi0\pi}$ phase transition. This suggests that, at least within our mean-field theory, the proximity of $c$-electron Fermi surfaces to the Dirac points favors an HW instability of the FL$^*_{0\pi0\pi}$ phase \cite{node_note}.

We note that a large-$N$ mean-field study on a related square-lattice model also encountered HW solutions near a KB transition to an FL$^*$ phase with Dirac spinons \cite{drechsler25}. Together with our results, this strengthens the case for a connection between both phenomena.

\paragraph{Perturbation theory---} To gain insight into the origin of HWs, we perform a perturbative analysis around the FL$^*$. As detailed in the End Matter (EM), this starts from the path-integral representation of a Kondo-Heisenberg lattice model, written under the assumption that the $c$ and $f$ sectors are invariant under translations of a two-site unit cell. Contrary to our mean-field theory, no further assumptions are made regarding the nature of the $b$ and $\chi$ fields, nor the range of the latter. Hence, the following conclusions apply to a wider class of fractionalized Fermi liquids than just the FL$^*_{0\pi0\pi}$ phase above.
By integrating out the $c$ and $f$ fermions, we derive an effective action $S_\mathrm{eff}[\lambda, \chi, b]$ as a power series in $J_K b$. Writing $S_\mathrm{eff} = S_{\lambda\chi} + S_b$, where $S_b$ encodes the full $b$ dependence, we obtain
\begin{align}
	S_b
	&= N J_K^2 \sum_{a, a',q} b^*_{a} (q) \left[
	\frac{\delta_{aa'}}{J_K} -
	\Pi_{aa'} (q)
	\right] b_{a'} (q)
	+ \mathcal{O} (b^4).
	\label{eq:Fb1}
\end{align}
Here, $a$ is a sublattice index, $q = (\vec{q}, \ii \nu_n)$ is formed by a wavevector $\vec{q}$ and bosonic Matsubara frequency $\nu_n$, and $\Pi_{aa'} (q)$ denotes a polarization bubble given in terms of the free $c$ and $f$ propagators [see Eq.~\eqref{eq:Pi}].

In the weak-coupling limit $J_K \to 0^+$, the positive term $\mathds{1}/J_K$ dominates the quadratic kernel in Eq.~\eqref{eq:Fb1}, thereby preventing the condensation of $b$ and ensuring the absence of Kondo screening. However, increasing $J_K$ gradually reduces the stiffness of the hybridization fields until Eq.~\eqref{eq:Fb1} becomes unstable at a critical coupling $J_K^c$. If not preempted by a first-order transition, this instability drives a continuous transition from the FL$^*$ to a Kondo-screened phase whose properties are determined by a free energy $F_b = \beta^{-1}S_b|_{\nu_n=0}$ corresponding to the \emph{static} part of $S_b$ \cite{suppl}. After diagonalizing $\Pi(\vec{q}) \equiv \Pi(\vec{q}, 0)$, we obtain
\begin{equation}
	F_b
	= \frac{N J_K^2}{\beta} \sum_{\vec{q}} \sum_{s=\pm}
	\left[ \frac{1}{J_K} - p_s (\vec{q}) \right]
	\abs{ b_s (\vec{q}) }^2
	+ \mathcal{O} (b^4),
	\label{eq:Fb2}
\end{equation}
with $b_s (\vec{q}) \equiv b_s (\vec{q}, 0)$ and
\begin{align}
	p_\pm &=
	\frac{\Pi_{11} + \Pi_{22}}{2} \pm
	\sqrt{ \left( \frac{\Pi_{11} - \Pi_{22}}{2} \right)^2 + \abs{\Pi_{12}}^2 }.
	\label{eq:ppm}
\end{align}
Thus, once $J_K \geq J_K^c = [\max_\vec{q} p_{+} (\vec{q})]^{-1}$, $b_+$ can condense at momenta $\vec{Q}$ that maximize $p_{+} (\vec{q})$. As discussed in the SM \cite{suppl}, the periodicity of the resulting Kondo-screened state is set not only by the values of such optimal wavevectors, but also by the number of $b_{+} (\vec{Q})$ modes that condense.
While gauge invariance implies that there can be no \emph{inter-unit-cell} modulations if $\expval{b_{+} (\vec{Q})} \ne 0$ for a single $\vec{Q}$, our theory allows for the simultaneous condensation of one or even two pairs of modes with opposite momenta $\pm\vec{Q}$ -- situations which give rise to single-$\vec{q}$ and multi-$\vec{q}$ HWs, respectively. Screening variations \emph{within} the two-site unit cell are nonetheless possible in all of the above scenarios and are dictated by sublattice-dependent weights, $w_A$ and $w_B$, with $w_a (\vec{Q}) = \expval{b_a (\vec{Q})} / \expval{ b_{+} (\vec{Q}) }$ given in Eq.~\eqref{eq:bsub}.

Overall, this framework captures a much wider range of instabilities than our mean-field theory, which is restricted to $\vec{Q}=\vec{0}$ due to the assumption that \emph{hybridized} states---and not just the isolated $c$ and $f$ sectors---are invariant under translations of the two-site unit cell.

The computation of $\Pi(\vec{q})$ yields a weighted sum of four Lindhard-like functions, each of which describes virtual particle-hole transitions between a different pair of $c$ and $f$ bands \cite{suppl}. Although generally cumbersome, the result simplifies considerably in the limit $T\to 0$ if one assumes that (i) the $c$-electron hopping Hamiltonian has $A\leftrightarrow B$ sublattice symmetry and that, like the DSL ansatz, (ii) the FL$^*$ hosts particle-hole-symmetric spinon bands $\pm \eps_{\vec{k}}^f$ with $\eps_{\vec{k}}^f \ge 0$. As discussed in the EM [see Eqs.~\eqref{eq:Pidiagsum}-\eqref{eq:Pioffdiag}], all three contributions $(\Pi_{11} + \Pi_{22})$, $(\Pi_{11} - \Pi_{22})$ and $\abs{\Pi_{12}}$ entering Eq.~\eqref{eq:ppm} are given in terms of momentum-space integrals of functions proportional to $( \eps_\vec{k}^f + | \xi_{\vec{k} + \vec{q}}^{cs} | )^{-1}$, where $\xi_{\vec{k}}^{cs}$ represent bare $c$-electron bands measured relative to the chemical potential. This suggests that, if $\eps_{\vec{k}}^f$ has nodes at Dirac points $\pm\vec{K}$, then the maxima of $p_{+}$ should arise at momenta $\vec{Q}$ such that $\xi_{\vec{K} + \vec{Q}}^{cs} \sim 0$ or $\xi_{-\vec{K} + \vec{Q}}^{cs} \sim 0$ for some band $s$ that participates in the bare $c$ Fermi surface.
For fixed $\vec{Q} = \vec{0}$, this implies that the $c$ Fermi surface(s) must be near the Dirac points, which is precisely the condition we inferred at the mean-field level.

\begin{figure}
	\centering
	\includegraphics[width=\columnwidth]{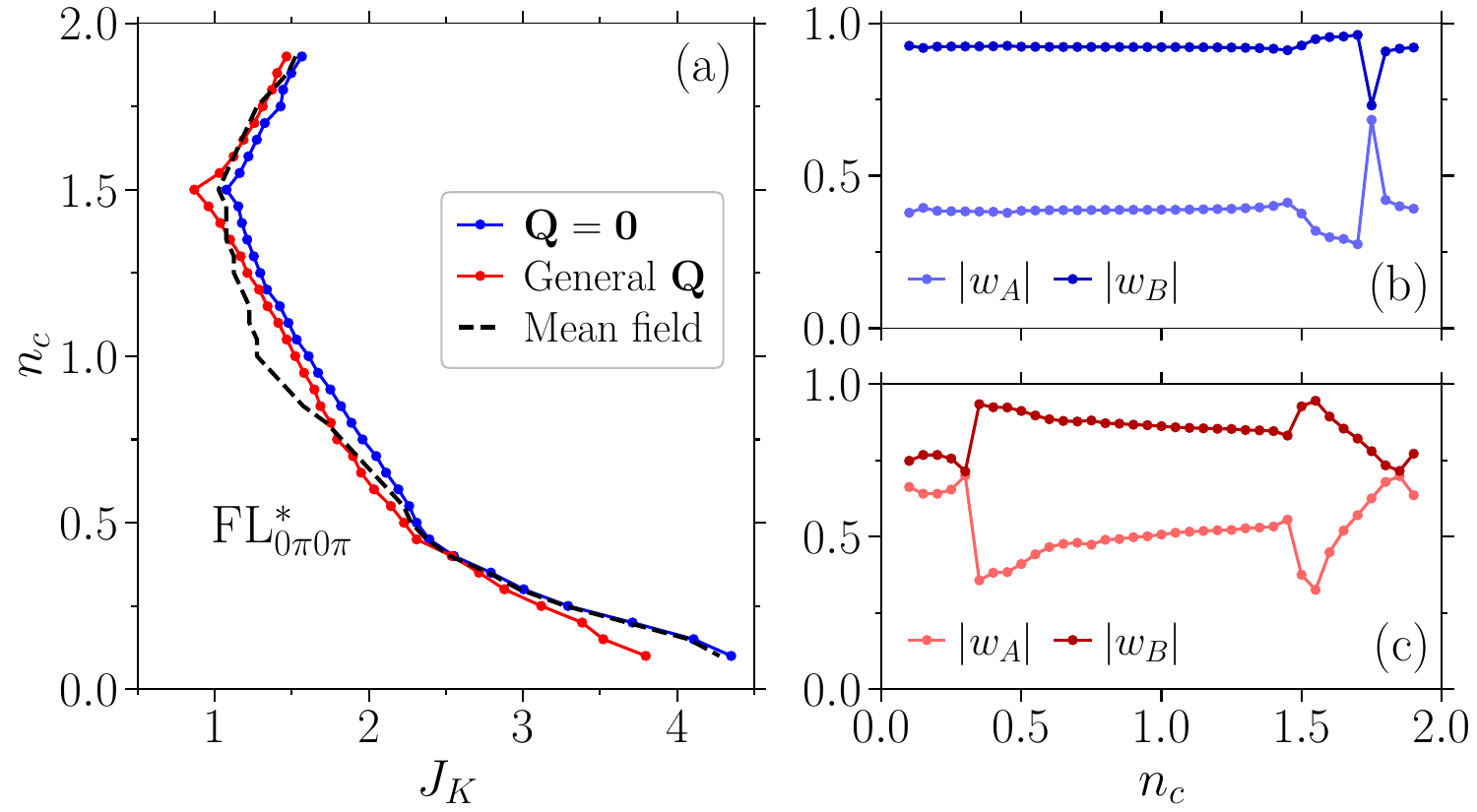}
	\caption{Perturbation-theory results for $L=300$, $J_H=0.2$, $\kappa=1$, and $k_B T = 10^{-3}$. (a) Comparison between the mean-field FL$^*_{0\pi0\pi}$ phase boundary from Fig.~\ref{fig:mfplots}(b) (black) and the critical couplings $J_K^c$ obtained for the restricted $\vec{Q}=0$ (blue) and the general-$\vec{Q}$ (red) theories \cite{systemsize_note}. (b,c) Sublattice weights $\abs{w_a (\vec{Q})} = \abs{\expval{b_a (\vec{Q})} / \expval{b_{+} (\vec{Q})}}$ for the onset of screening in (b) $\vec{Q}=0$ and (c) general-$\vec{Q}$ instabilities.}
	\label{fig:ptgrid}
\end{figure}

Figure~\ref{fig:ptgrid} shows perturbative results derived with the same dispersions $\xi_{\vec{k}s}^c$ and $\eps_{\vec{k}}^f$ used in our mean-field calculations. In Fig.~\ref{fig:ptgrid}(a), we compare the FL$^*_{0\pi0\pi}$ phase boundary from Fig.~\ref{fig:mfplots}(b) with the critical Kondo coupling, $J_K^c (n_c)$, obtained both for the cases of a general instability wavevector $\vec{Q}$ and a theory restricted to $\vec{Q}=0$. As expected, the $\vec{Q}=0$ and mean-field results agree well when there is a continuous FL$^*_{0\pi0\pi}$--HW$_{0\pi0\pi}$ transition. However, they only both match the general-$\vec{Q}$ results around $n_c\approx 0.45$. For generic $n_c$,  we either find that $\vec{Q} \ne 0$ instabilities take precedence and lead to lower $J_K^c$ or that the system exits the FL$^*$ through a first-order transition.
Meanwhile, Figs.~\ref{fig:ptgrid}(b,c) illustrate how the sublattice-dependent weights $w_a$ vary as a function of $n_c$ for (b) $\vec{Q}=0$, where this is the only possible source of HWs, and (c) generic $\vec{Q}$. Both cases show that inhomogeneities within the unit cell are practically ubiquitous, such that a continuous KB transition \emph{generally} involves a state with modulated Kondo screening. In the EM, we show that the soft $\vec{Q}$ wavevectors indeed satisfy the aforementioned condition $\xi_{\pm\vec{K} + \vec{Q}}^{cs} \sim 0$.


\paragraph{Summary and outlook---} 
We have shown that, under the right kinematic conditions, heavy-fermion metals in proximity to an FL$^*$ phase with Dirac spinons develop HWs. Together with previous work from Refs.~\onlinecite{paul07,zhu08,drechsler25}, this result establishes a strong link between HWs and KB transitions involving gapless fermionic spinons. Importantly, our perturbatively derived field theory goes far beyond previous lattice mean-field calculations in that it captures the conditions for generic HW instabilities and orders.

Future work should consider whether the onset of HWs can lead to a simultaneous change in the U(1) fluxes $\Phi$ and thereby give rise to a chiral metal that breaks time-reversal symmetry \cite{drechsler25}. This effect would result from a coupling between the $b$ and $\chi$ hybridization fields in the effective free energy and leave distinctive signatures in electronic spectral functions probed by angle-resolved photoemission spectroscopy.
More generally, the quantum critical behavior at the KB transition between FL$^\ast$ and HW phase---which now involves both symmetry breaking and confinement---should be studied in detail.
Another potential avenue for investigation consists of alternative mechanisms for the emergence of Kondo HWs that do not rely on fractionalization or frustration.


\paragraph{Note added---} During the preparation of our manuscript, we became aware of a preprint \cite{zhong26} that demonstrates the emergence of HWs in a large-$N$ mean-field theory of a Kondo lattice model. This work restricts itself to states invariant under translations of a two-site unit cell and does not incorporate interactions between the local moments.


\paragraph{Acknowledgements---} We thank Tom Drechsler, Yasir Iqbal, Shi-Zeng Lin, Daniel Lozano-G\'omez, and Urban F. P. Seifert for illuminating discussions and for collaborations on related topics.
Part of this work (mean-field theory, perturbative analysis) was supported by the U.S. Department of Energy, Office of Science, Office of Basic Energy Sciences, Material Sciences and Engineering Division under Award Number DE-SC0025247 (PMC and OE).
MV acknowledges support by the DFG through SFB 1143 (project id 247310070) and the W\"urzburg-Dresden Cluster of Excellence on Complexity, Topology and Dynamics in Quantum Matter---\textit{ctd.qmat} (EXC 2147, project id 390858490).
We also acknowledge Research Computing at Arizona State University for providing HPC resources \cite{sol} that contributed to the results reported in this paper.


\bibliography{kondotr}


\appendix

\newpage
\onecolumngrid
\section*{End Matter}
\twocolumngrid

\paragraph{Details on the perturbative calculation---} The perturbative calculation outlined in the main text starts with the coherent-state path integral representation of the grand-canonical partition function of a Kondo-Heisenberg lattice model \cite{senthil04,coleman_book}. Assuming that the pure $c$ and $f$ sectors are invariant under translations of a two-site unit cell, we divide the action $S = S_0 + S_b^0 + S_b^\mathrm{int}$ into three parts:
\begin{subequations}
	\begin{align}
		S_0 &=
		\int_0^\beta d\tau \left[ \sum_{\vec{k}\sigma} \sum_{a,a'}
		c_{\vec{k}a\sigma}^* \left( \delta_{aa'} \partial_\tau + h_{c\vec{k}}^{aa'} \right) c_{\vec{k}a'\sigma}
		\right.
		\notag \\
		&
		\phantom{=\int_0^\beta d\tau}
		+\sum_{\vec{k}\sigma} \sum_{a,a'}
		f_{\vec{k}a\sigma}^* \left( \delta_{aa'} \partial_\tau + h_{f\vec{k}}^{aa'} \right) f_{\vec{k}a'\sigma}
		\notag \\
		&
		\phantom{=\int_0^\beta d\tau} \left.
		+\mathcal{L}_\chi - \frac{N}{2} \sum_{ia} \lambda_{ia} \right],
		\label{eq:S0}
		\\
		S_b^0 &= NJ_K \int_0^\beta d\tau \sum_{ia} b_{ia}^* b_{ia},
		\label{eq:Sb0}
		\\
		S_b^\mathrm{int} &= -J_K \int_0^\beta d\tau \sum_{ia\sigma} \left( b_{ia} c_{ia\sigma}^* f_{ia\sigma} + \mathrm{c.c.} \right),
		\label{eq:Sbint}
	\end{align}
\end{subequations}
where the Lagrangian $\mathcal{L}_\chi$ encodes the dependence on the $\chi$ fields and $\kappa$ and $N$ denotes the number of fermionic flavors $\sigma$. Furthermore, $h_{c\vec{k}}$ and $h_{f\vec{k}}$ are $2\times 2$ matrices that reflect the sublattice content of the theory. Contrary to $h_{c\vec{k}}$, $h_{f\vec{k}}$ depends on the $\chi_{jb,ia}$ and $\lambda_{ia} \equiv \lambda_a$ fields, which are not constrained by any conditions besides translation invariance.
When explicit calculations are involved, we shall consider
\begin{align}
	h_{c\vec{k}} &= -2t
	\begin{pmatrix}
		\cos k_1 & \cos k_2 + \cos k_3
		\\
		\cos k_2 + \cos k_3 & \cos k_1
	\end{pmatrix} - \mu \mathds{1},
	\label{eq:hkc}
	\\
	h_{f\vec{k}} &= -2J_H \chit
	\begin{pmatrix}
		\cos k_1 & \cos k_2 + \ii \sin k_3
		\\
		\cos k_2 - \ii \sin k_3 & -\cos k_1
	\end{pmatrix}.
	\label{eq:hkf}
\end{align}
Here, $\mu$ is the chemical potential, $\mathds{1}$ represents the $2\times 2$ identity matrix, and $k_n = \vec{k} \cdot \bm{\delta}_n$ denotes the projection of $\vec{k}$ along the nearest-neighbor vectors $\bm{\delta}_1 = (1,0)$, $\bm{\delta}_2 = \frac{1}{2} (1, \sqrt{3})$ and $\bm{\delta}_3 = \frac{1}{2}(-1, \sqrt{3})$. The matrices in Eqs.~\eqref{eq:hkc} and \eqref{eq:hkf} follow from the assumption of uniform nearest-neighbor hopping for the $c$ electrons and the DSL ansatz described in the SM \cite{suppl}.

We can derive an effective action $S_\mathrm{eff} = S_{\lambda\chi} + S_b$ that only depends on the bosonic fields by integrating out the fermions in a power series in $J_K b$. Up to second order, the result is given by
\begin{align}
	e^{-S_\mathrm{eff}} &= \int \mathcal{D} (c^*, c) \mathcal{D} (f^*, f) \; e^{-S}
	\notag \\
	&\approx
	e^{-S_{\lambda\chi}} \exp \left[- S_b^0 + \frac{1}{2} \expval{ \left( S_b^\mathrm{int} \right)^2 }_0 \right],
	\label{eq:integrateout}
\end{align}
where $\expval{\cdots}_0$ denotes an average with respect to $S_0$. Equation~\eqref{eq:Fb1} is then obtained by using Wick's theorem to evaluate the term $\expval{ \left( S_b^\mathrm{int} \right)^2 }_0$ in Eq.~\eqref{eq:integrateout}.

\begin{figure*}[t]
	\centering
	\includegraphics[width=\textwidth]{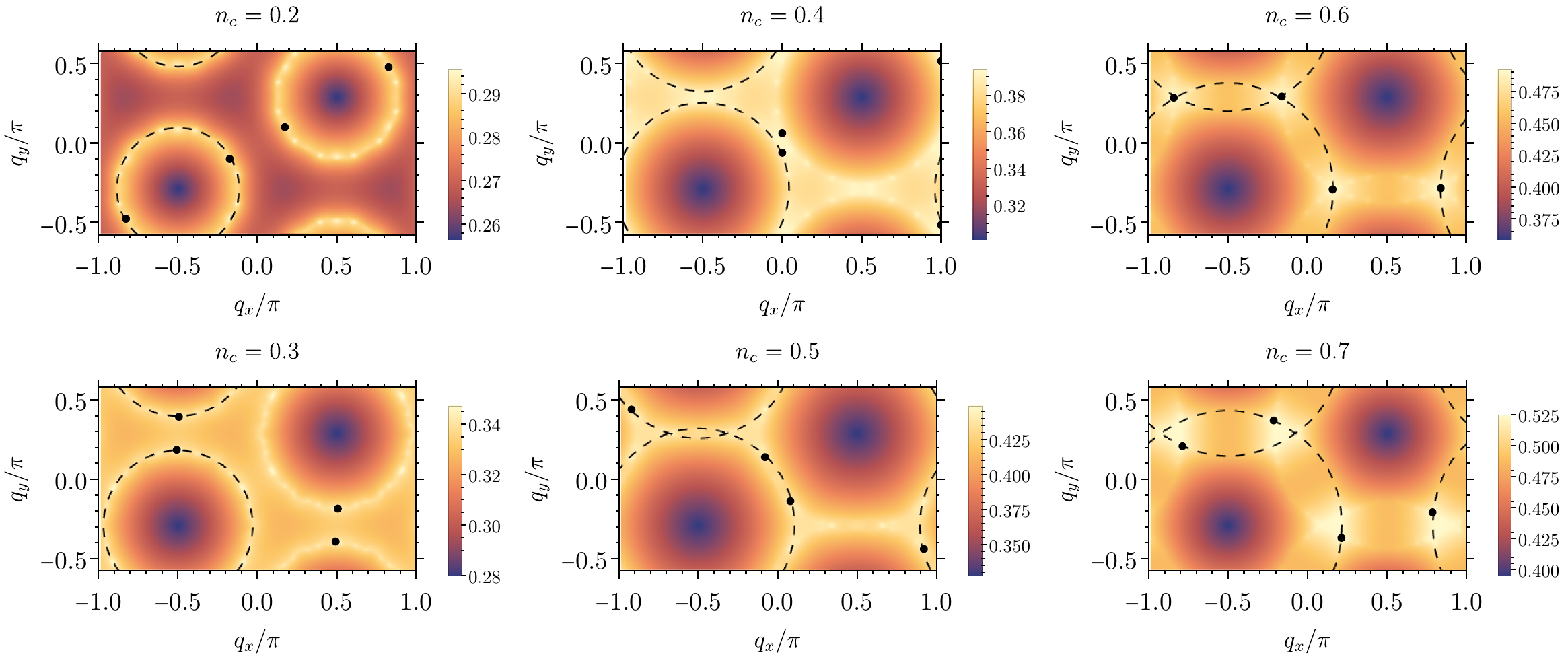}
	\caption{Color maps of the function $p_{+} (\vec{q})$ for different fillings. The data were generated for a finite system with $L = 150$ and parameters $k_BT=10^{-3}$, $J_H = 0.2$, and $\kappa=1$. Each plot contains four black dots indicating the positions of the maxima $\vec{Q}$, which determine the critical Kondo coupling, $J_K^c = \left[ \max_\vec{q} p_{+} (\vec{q}) \right]^{-1}$. The dashed lines correspond to the contours $\xi_{\vec{K} + \vec{q}}^{c\pm} = 0$, with $\vec{K} = (\frac{\pi}{2}, \frac{\pi}{2\sqrt{3}})$ being one of the nodes in the dispersion of the $f$ spinons.}
	\label{fig:pmqgrid}
\end{figure*}

The next step is to compute the polarization bubble
\begin{align}
	-\Pi_{aa'} (q) &=
	\;\;
	\begin{tikzpicture}[baseline=(current bounding box.center)]
		\begin{feynman}
			\vertex [dot] (a);
			\vertex [dot, right = of a] (b);
			\diagram*{
				(a) -- [fermion, half left, looseness=1.2, edge label = $k+q$] (b),
				(b) -- [charged scalar, half left, looseness=1.2, edge label = $k$] (a)
			};
			\node[dot, label=left:$a'$] at (a) {};
			\node[dot, label=right:$a$] at (b) {};
		\end{feynman}
	\end{tikzpicture}
	\;\;
	\notag \\
	&= \frac{1}{\beta N_c} \sum_k G^{aa'}_{c0} (k+q) G^{a'a}_{f0} (k),
	\label{eq:Pi}
\end{align}
where $N_c = N_s/2$ and $G^{aa'}_{\ell 0} (k) = -\expval{\ell_{a\sigma}(k) \ell^*_{a'\sigma} (k)}_0$ are free fermionic propagators. After evaluating the sum over Matsubara frequencies in Eq.~\eqref{eq:Pi}, one obtains an explicit expression for $\Pi (q)$ in terms of the eigenvalues $\xi_{\vec{k}}^{\ell s}$ of the matrices $h_{\ell\vec{k}}$, with $s = \pm$ \cite{suppl}.

The instability of the FL$^*$ to the condensation of the hybridization fields and to the ensuing emergence of a Kondo-screening phase is determined by the static part of the polarization bubble, $\Pi(\vec{q}) \equiv \Pi(\vec{q},0)$. Taking the limit $T\to 0$ and assuming only that (i) $h_{c\vec{k}}$ has the full periodicity of the underlying triangular lattice and (ii) $\xi_{\vec{k}}^{f\pm} = \pm \eps_{\vec{k}}^f$ with $\eps_{\vec{k}}^f \ge 0$, we obtain \cite{suppl}
\begin{subequations}
\begin{align}
	&\frac{\Pi_{11} + \Pi_{22}}{2} = \frac{1}{4N_c} \sum_{\vec{k}s} \frac{1}{\eps_\vec{k}^f + \abs{\xi_{\vec{k}+\vec{q}}^{cs}}},
	\label{eq:Pidiagsum}
	\\
	&\frac{\Pi_{11} - \Pi_{22}}{2} = \frac{1}{4N_c} \sum_{\vec{k}s}
	\frac{h_{f\vec{k}}^{11}}{\eps_\vec{k}^f} \frac{
		\left[
		1  - 2\Theta\left( \xi_{\vec{k} + \vec{q}}^{cs} \right)
		\right]
	}{
		\left( \eps_\vec{k}^f + \abs{\xi_{\vec{k}+\vec{q}}^{cs}} \right)
	},
	\label{eq:Pidiagdiff}
	\\
	&\Pi_{12} = \frac{1}{4N_c} \sum_{\vec{k}s}
	s \frac{h_{c,\vec{k}+\vec{q}}^{12}}{\abs{h_{c,\vec{k}+\vec{q}}^{12}}} \frac{h_{f\vec{k}}^{21}}{\eps_\vec{k}^f} \frac{
		\left[
		1  - 2\Theta\left( \xi_{\vec{k} + \vec{q}}^{cs} \right)
		\right]
	}{\left( \eps_\vec{k}^f + \abs{\xi_{\vec{k}+\vec{q}}^{cs}} \right)},
	\label{eq:Pioffdiag}
\end{align}
\end{subequations}
where $\Theta$ is the Heaviside function. Note that the presence of $\eps_{\vec{k}}^f$ in the denominators of the summands of Eqs.~\eqref{eq:Pidiagdiff} and \eqref{eq:Pioffdiag} does not necessarily lead to (integrable) singularities at the Dirac points $\vec{k} = \pm\vec{K}$. Indeed, one can check that the magnitudes of $(h_{f\vec{k}}^{11} / \eps_{\vec{k}}^f)$ and $(h_{f\vec{k}}^{21} / \eps_{\vec{k}}^f)$ are bounded near $\pm\vec{K}$ when $h_{f\vec{k}}$ is given in Eq.~\eqref{eq:hkf}. On the other hand, the factors $( \eps_\vec{k}^f + | \xi_{\vec{k} + \vec{q}}^{cs} | )^{-1}$ indicate that all of the above terms -- and consequently the function $p_{+}(\vec{q})$ -- are enhanced for momenta $\vec{Q}$ such that $\xi_{\pm\vec{K} + \vec{Q}}^{cs} \sim 0$.

As discussed in the main text, one can determine the critical Kondo coupling $J_K^c$ at a fixed filling $n_c$ by maximizing $p_{+}(\vec{q})$, the highest eigenvalue of $\Pi(\vec{q})$. The results in Fig.~\ref{fig:pmqgrid} demonstrate that this function generally has \emph{four} degenerate global maxima at momenta $\pm\vec{Q}_1$ and $\pm \vec{Q}_2$. In the SM \cite{suppl}, we show that this fourfold degeneracy is imposed by the invariance of $S$ under two (projective) symmetries associated with (i) time reversal and (ii) spatial translations that exchange sublattices $A$ and $B$ [see Fig.~\ref{fig:mfplots}(a)]. Whereas (i) enforces that $p_{+}( \vec{q} )$ is even in $\vec{q}$, (ii) implies that $p_{+}(\vec{q}_1) = p_{+}(\vec{q}_2)$ for momenta $\vec{q}_2 = \vec{q}_1 + \vec{b}_1/2$ connected by half of a reciprocal lattice vector $\vec{b}_1 = (2\pi, -2\pi/\sqrt{3})$. In particular, this latter property guarantees that $p_{+}(\vec{q})$ has half of the periodicity of the reduced Brillouin zone. Another interesting feature illustrated in Fig.~\ref{fig:pmqgrid} is that the contours $\xi_{\vec{K} + \vec{Q}}^{c\pm}$ (black dashed lines) generally cross (or very nearly cross) maxima of $p_{+}$. The outlier maxima seen in the cases $n_c < 0.5$ can be shown to intersect the contour $\xi_{-\vec{K} + \vec{Q}}^{c\pm}$, which were omitted for clarity. Thus, our data fully support the earlier claim that the possible condensation wavevectors obey $\xi_{\pm\vec{K} + \vec{Q}}^{cs} \sim 0$.

The diagonalization of $\Pi(\vec{q})$ also establishes the relation between the expectation values of the original hybridization fields, $b_A$ and $b_B$, in Eq.~\eqref{eq:Fb1} and the field $b_{+}$ that condenses in Eq.~\eqref{eq:Fb2}. From the result
\begin{align}
	\begin{pmatrix}
		\expval{b_A} \\
		\expval{b_B}
	\end{pmatrix}
	&=
	\frac{ \expval{b_{+}} }{\sqrt{ (p_{+} - \Pi_{11})^2 + \abs{\Pi_{12}}^2 }}
	\begin{pmatrix}
		\Pi_{12} \\
		p_{+} - \Pi_{11}
	\end{pmatrix},
	\label{eq:bsub}
\end{align}
one can read off the sublattice weights $w_a(\vec{Q}) = \expval{b_a (\vec{Q})} / \expval{ b_{+} (\vec{Q}) }$, whose absolute values are shown in Fig.~\ref{fig:ptgrid}(b,c). Notably, we find that $w_a (-\vec{Q}) = w_a(\vec{Q})^*$ and $\abs{w_A (\vec{Q} \pm \vec{b}_1/2) } = \abs{w_B (\vec{Q})}$. These properties are consistent with the interpretation that the pairs of maxima $\pm\vec{Q}_1$ and $\pm\vec{Q}_2$ of $p_{+}$ are related to each other by exchanging the sublattices $A\leftrightarrow B$.


\end{document}


\title{Supplemental material for: ``Heavy-fermion hybridization waves from Dirac spinons near a Kondo-breakdown quantum phase transition''}
\author{Pedro M. C\^onsoli}
\affiliation{Department of Physics, Arizona State University, Tempe, AZ 85287, USA}
\author{Matthias Vojta}
\affiliation{Institut f\"ur Theoretische Physik and W\"urzburg-Dresden Cluster of Excellence ctd.qmat, Technische Universit\"at Dresden, 01062 Dresden, Germany}
\author{Onur Erten}
\affiliation{Department of Physics, Arizona State University, Tempe, AZ 85287, USA}

\date{\today}
\maketitle
\tableofcontents


\section{Large-$N$ mean-field theory of the Kondo-Heisenberg lattice model}
\label{sec:mft}

In this section, we provide details on the large-$N$ mean-field theory discussed in the main text. We start by writing the SU($N$)-symmetric generalization of the Hamiltonian in Eq.~(1),
\begin{equation}
	\HH =
	- \sum_{\langle ij \rangle, \alpha} \left( t_{ij} c_{i\alpha}^\dagger c_{j\alpha} + \hc \right)
	+ \frac{J_K}{N} \sum_{i,\alpha,\beta} S_{i}^{\beta\alpha} c_{i\alpha}^\dagger c_{i\beta}
	+ \frac{J_H}{N} \sum_{\langle ij \rangle} \left[ \sum_{\alpha\beta} S_{i}^{\beta\alpha} S_{j}^{\alpha\beta} + \frac{\kappa}{N^2} \left( \sum_{\alpha\beta} S_{i}^{\beta\alpha} S_{j}^{\alpha\beta}\right)^2 \right],
	\label{eq:khlm_sum}
\end{equation}
where $\alpha,\beta = 1,\ldots,N$ and $S_i^{\beta\alpha}$ are generators of the fundamental representation of SU($N$). The generators are then represented in terms of complex fermions via the relation
\begin{equation}
	S_i^{\beta\alpha} = f_{i\beta}^\dagger f_{i\alpha} - \frac{1}{2} \delta_{\alpha\beta}
	\label{eq:partonrep}
\end{equation}
and the associated constraint
\begin{equation}
	\sum_{\alpha=1}^N f_{i\alpha}^\dagger f_{i\alpha} = \frac{N}{2}.
	\label{eq:fconstraint}
\end{equation}
The usual Abrikosov representation is recovered for $N=2$ if one uses the relations $S_i^z = (S_i^{11} - S_i^{22})/2$ and $S_i^+ = S_i^{12}$.

Given the choice of the nontrivial mean-field unit cell shown in Fig.~2(a), we can label the sites of the triangular lattice in terms of a composite index $(ia)$, where $i$ numbers one of the $N_c$ unit cells in the system and $a \in \{ A\equiv1, B\equiv2 \}$ is a sublattice index. We denote the total number of sites as $N_s = 2N_c$. Working in units where the lattice spacing is one, we also define the nearest-neighbor vectors $\bm{\delta}_1 = (1,0)$, $\bm{\delta}_2 = (1,\sqrt{3})/2$, and $\bm{\delta}_3 = (-1,\sqrt{3})/2$. With these conventions, the $c$-electron hopping term can be written as 
\begin{align}
	\HH_t &= -2t \sum_{\vec{k}\sigma} \left[
	\cos k_1 \sum_{a} c_{\vec{k} a\sigma}^\dagger c_{\vec{k} a\sigma} + \left( \sum_{n=2}^{3} \cos k_n \right) \left( c_{\vec{k} A \sigma}^\dagger c_{\vec{k} B \sigma} + \hc \right)
	\right],
	\label{eq:Ht}
\end{align}
where $c_{\vec{k}a\sigma} = N_c^{-1/2} \sum_i e^{i\vec{k} \cdot \vec{r}_{ia}} c_{ia\sigma}$ and $k_n = \vec{k} \cdot \bm{\delta}_n$. Defining the mean fields
\begin{align}
	b_a &= \frac{1}{N} \sum_{\sigma} \expval{ f_{ia\sigma}^\dagger c_{ia\sigma} },
	&
	\chi_{jb, ia} &= \frac{1}{N} \sum_{\sigma} \expval{ f_{jb\sigma}^\dagger f_{ia\sigma} } = \chi_{ia, jb}^*,
\end{align}
we decouple the Kondo and Heisenberg interactions according to
\begin{align}
	\sum_{\alpha\beta} S_{ia}^{\beta\alpha} c_{ia\alpha}^\dagger c_{ia\beta} &= 
	 \frac{N}{2} - \frac{1}{2} \sum_\alpha c_{ia\alpha}^\dagger c_{ia\alpha} - \sum_{\alpha\beta} f_{ia\beta}^\dagger c_{ia\beta} c_{ia\alpha}^\dagger f_{ia\alpha}
	\notag \\
	&\approx 
	\frac{N}{2} - \frac{1}{2} \sum_\sigma c_{ia\sigma}^\dagger c_{ia\sigma} - N \sum_\sigma \left( b_a c_{ia\sigma}^\dagger f_{ia\sigma} + \hc \right) + N^2 \abs{b_a}^2,
	\\
	\sum_{\alpha\beta} S_{ia}^{\beta\alpha} S_{jb}^{\alpha\beta} &= \frac{N}{4} - \sum_{\alpha\beta} f_{ia\beta}^\dagger f_{jb\beta} f_{jb\alpha}^\dagger f_{ia\alpha}
	\notag \\
	&\approx
	\frac{N}{4} - N \sum_\sigma \left( \chi_{jb, ia} f_{ia\sigma}^\dagger f_{jb\sigma} + \hc \right) + N^2 \abs{\chi_{jb, ia}}^2.
	\label{eq:decoupheis}
\end{align}
The biquadratic interaction is then treated by squaring Eq.~\eqref{eq:decoupheis} and applying the same large-$N$ decoupling to the remaining quartic contribution \cite{drechsler25}. To leading order in $N$, this yields:
\begin{align}
	\left( \frac{1}{N} \sum_{\alpha\beta} S_{ia}^{\beta\alpha} S_{jb}^{\alpha\beta} \right)^2
	&\approx
	- 3N^2 \abs{\chi_{jb,ia}}^4 
	+ 2N \abs{ \chi_{jb,ia} }^2 \sum_\sigma \left( \chi_{jb,ia} f_{i\sigma}^\dagger f_{j\sigma} + \hc \right).
\end{align}
Finally, we introduce static, sublattice-dependent Lagrange multipliers $\lambda_a$ to enforce that the constraints in Eq.~\eqref{eq:fconstraint} are satisfied \emph{on average}. With this, we obtain the large-$N$ mean-field Hamiltonian
\begin{align}
	\HH_\mathrm{MF} 
	&= \HH_t + \sum_{ia} \lambda_a \left( f_{ia\sigma}^\dagger f_{ia\sigma} - \frac{N}{2} \right)
	- J_K \sum_{ia} \left( b_a c_{ia\sigma}^\dagger f_{ia\sigma} + \hc - N \abs{b_a}^2 \right)
	\notag \\
	&- J_H \sum_{\langle ia, jb \rangle} \left[ \tilde{\chi}_{jb, ia} f_{ia\sigma}^\dagger f_{jb\sigma} + \hc 
	+ N \abs{\chi_{jb, ia}}^2 \left( 3\kappa \abs{\chi_{jb, ia}}^2 - 1 \right)
	\right],
	\label{eq:hmf1}
\end{align}
where sums over repeated flavor indices are implied and
\begin{equation}
	\tilde{\chi}_{jb, ia} = \left( 1 - 2\kappa \abs{\chi_{jb, ia}}^2 \right) \chi_{jb, ia}.
\end{equation}

Given that the unit cell depicted in Fig.~2(a) accommodates six nearest-neighbor bonds, it admits six independent complex fields $\chi_{an} \equiv \chi_{(ia)+\bm{\delta}_n, (ia)}$. Two of these, $\chi_{A1}$ and $\chi_{B1}$, connect sites on the same sublattice. After performing a Fourier transform, we can write the grand-canonical operator $\KK_\mathrm{MF} = \HH_\mathrm{MF} - \mu \mathcal{N}_c$, where $\mathcal{N}_c = \sum_{ia\sigma} c_{ia\sigma}^\dagger c_{ia\sigma}$, as
\begin{equation}
	\KK_\mathrm{MF} = 
	E_0 + \sum_{\vec{k}\sigma} \Psi_{\vec{k}\sigma}^\dagger
	\underbrace{ 
		\begin{pmatrix}
			t G (\vec{k}) - \mu \mathds{1}_2 & -J_K B
			\\
			-J_K B^* & J_H G_\chi (\vec{k}) + \Lambda
		\end{pmatrix}
	}_{=h_\vec{k}}
	\Psi_{\vec{k}\sigma},
	\label{eq:kmf}
\end{equation}
with $\Psi_{\vec{k}\sigma}^\top = \begin{pmatrix} c_{\vec{k}A\sigma} & c_{\vec{k}B\sigma} & f_{\vec{k}A\sigma} & f_{\vec{k}B\sigma} \end{pmatrix}$ and
\begin{equation}
	E_0 = NN_c \left[
	\sum_a \left( J_K \abs{b_a}^2 - \frac{\lambda_a}{2} \right) + J_H \sum_{an} \abs{\chi_{an}}^2 \left( 1 - 3\kappa \abs{\chi_{an}}^2 \right)
	\right].
	\label{eq:e0}
\end{equation}
The $4\times 4$ matrix $h_\vec{k}$ is defined in terms of the $2\times 2$ submatrices $B_{ab} = b_a \delta_{ab}$, $\Lambda_{ab} = \lambda_a \delta_{ab}$, and
\begin{align}
	G(\vec{k}) &= -2
	\begin{pmatrix}
		\cos k_1 & \cos k_2 + \cos k_3
		\\
		\cos k_2 + \cos k_3 & \cos k_1
	\end{pmatrix},
	\\
	G_\chi (\vec{k}) &= -
	\begin{pmatrix}
		2\Re \left( e^{\ii k_1} \tilde{\chi}_{A1} \right) & \sum_{n=2}^3 \left( e^{\ii k_n} \tilde{\chi}_{An} + e^{-\ii k_n} \tilde{\chi}_{Bn}^* \right)
		\\
		\sum_{n=2}^3 \left( e^{\ii k_n} \tilde{\chi}_{Bn} + e^{-\ii k_n} \tilde{\chi}_{An}^* \right) & 2\Re \left( e^{\ii k_1} \tilde{\chi}_{B1} \right)
	\end{pmatrix}.
\end{align}
While the fields $b_a$ are in principle complex, we can eliminate their imaginary parts by performing a sublattice-dependent  gauge transformation $f_{\vec{k} a \sigma} \mapsto e^{-\ii \varphi_a} f_{\vec{k} a \sigma}$ with $\varphi_a = \arg(b_a)$.

As usual, Eq.~\eqref{eq:kmf} can be diagonalized via $\vec{k}$-dependent unitary transformations
\begin{align}
	\KK_\mathrm{MF} = E_0 + \sum_{\vec{k}\sigma} 
	\Psi_{\vec{k}\sigma}^\dagger U_\vec{k} 
	U_\vec{k}^\dagger h_\vec{k} U_\vec{k}
	U_\vec{k}^\dagger \Psi_{\vec{k}\sigma}
	\equiv
	E_0 + \sum_{\vec{k}\sigma} \sum_{n = 1}^{4} E_{\vec{k}n} \phi_{\vec{k}n\sigma}^\dagger \phi_{\vec{k}n\sigma},
\end{align}
where $E_{\vec{k}n}$ are the eigenvalues of $h_\vec{k}$ and $U_\vec{k}$ is the unitary matrix whose columns correspond to the eigenvectors of $h_\vec{k}$. The original fermionic operators are related to the new $\phi$ operators via the basis transformation $\Psi_{\vec{k}\sigma,m} = U_\vec{k}^{mn} \phi_{\vec{k}n\sigma}$. If we fix the $c$-electron filling $n_c = \expval{\mathcal{N}_c} / N_s$, then we have a total of eleven self-consistency equations:
\begin{subequations}
	\begin{align}
		n_c &= \frac{1}{N_s} \sum_{\vec{k}a\sigma} \expval{ c_{\vec{k}a\sigma}^\dagger c_{\vec{k}a\sigma} }
		&
		&\implies
		&
		n_c &= \frac{N}{N_s} \sum_{\vec{k}} \sum_{a=1}^{2} \sum_{n=1}^{4} \abs{U_\vec{k}^{an}}^2 n_F (E_{\vec{k}n}),
		\label{eq:mf1}
		\\
		\frac{N}{2} &= \frac{1}{N_c} \sum_{\vec{k}\sigma} \expval{ f_{\vec{k}a\sigma}^\dagger f_{\vec{k}a\sigma} }
		&
		&\implies
		&
		\frac{1}{2} &= \frac{1}{N_c} \sum_{\vec{k}} \sum_{n=1}^{4} \abs{ U_\vec{k}^{2+a, n} }^2 n_F (E_{\vec{k}n}),
		\label{eq:mf2}
		\\
		b_a &= \frac{1}{N N_c} \sum_{\vec{k} \sigma} \expval{f_{\vec{k}a\sigma}^\dagger c_{\vec{k}a\sigma}}
		&
		&\implies
		&
		b_a &= \frac{1}{N_c} \sum_{\vec{k}} \sum_{n=1}^{4} \left( U_\vec{k}^{2+a, n} \right)^* U_\vec{k}^{an} n_F (E_{\vec{k}n}),
		\label{eq:mf3}
		\\
		\chi_{a1} &= \frac{1}{N N_c} \sum_{\vec{k} \sigma} e^{-\ii k_1} \expval{ f_{\vec{k}a\sigma}^\dagger f_{\vec{k}a\sigma} }
		&
		&\implies
		&
		\chi_{a1} &= \frac{1}{N_c} \sum_{\vec{k}} \sum_{n=1}^{4} e^{-\ii k_1} \abs{ U_\vec{k}^{2+a, n} }^2 n_F (E_{\vec{k}n}),
		\label{eq:mf4}
		\\
		\chi_{a\ell} &= \frac{1}{N N_c} \sum_{\vec{k} \sigma} e^{-\ii k_\ell} \expval{ f_{\vec{k} \bar{a} \sigma}^\dagger f_{\vec{k} a \sigma} }
		&
		&\implies
		&
		\chi_{a\ell} &= \frac{1}{N_c} \sum_{\vec{k}} \sum_{n=1}^{4} e^{-\ii k_\ell} \left( U_\vec{k}^{2+\bar{a}, n} \right)^* U_\vec{k}^{2+a, n} n_F (E_{\vec{k}n}),
		\label{eq:mf5}
	\end{align}
\end{subequations}
where $n_F(x) = (e^{x/T} + 1)^{-1}$ is the Fermi-Dirac distribution function at temperature $T$. In Eq.~\eqref{eq:mf5}, $\ell=2,3$ and $\bar{a}$ represents the sublattice opposite to $a$, i.e., $\bar{a} = B$ if $a = A$ and vice-versa.

While the distinction between two mean-field phases may arise from differences in the hybridization fields $b_a$, it may also be expressed in gauge-invariant quantities obtained from the spinon hopping fields. As discussed in the main text, these quantities consist of the absolute values $\abs{\chi_{an}}$ of the bond variables and the four inequivalent fluxes
\begin{align}
	\Phi_1 &= \arg \left( \chi_{A2} \chi_{A3}^* \chi_{A1}^* \right),
	&
	\Phi_3 &= \arg \left( \chi_{B2} \chi_{B3}^* \chi_{B1}^* \right),
	\notag \\
	\Phi_2 &= \arg \left( \chi_{B1} \chi_{A2}^* \chi_{A3} \right),
	&
	\Phi_4 &= \arg \left( \chi_{A1} \chi_{B2}^* \chi_{B3} \right),
	\label{eq:fluxes}
\end{align}
depicted in Fig.~2(a). Note that the fluxes are defined in a clockwise orientation for every plaquette. Furthermore, $\exp (\ii \sum_{n=1}^4 \Phi_n) = 1$ implies that only three of the fluxes are independent.

Finally, in order to identify the ground states among multiple coexisting mean-field solutions, we must compute the (Helmholtz) free energy:
\begin{align}
	F &= -T \ln (\Tr e^{-\beta \HH_\mathrm{MF}} ) 
	= -T \ln (\Tr e^{-\beta \KK_\mathrm{MF}} ) + N_s \mu n_c
	\notag \\
	&= E_0 + N_s \mu n_c -TN \sum_{\vec{k}}\sum_{n=1}^{4} \ln \left( 1 + e^{-\beta E_{\vec{k}n}} \right).
\end{align}

\subsection{Pure $f$ sector: Unrestricted mean-field theory}

Having set up our parton mean-field theory, we now turn to a discussion of the solutions within the pure $f$ sector as a function of the biquadratic interaction $\kappa$.
An unrestricted numerical search for self-consistent solutions of Eqs.~\eqref{eq:mf2}, \eqref{eq:mf4}, and \eqref{eq:mf5} with $b_A = b_B = 0$ reveals that the mean-field ground state within the pure $f$ sector exhibits strong dimerization for $\kappa=0$, a trend previously observed on the square lattice \cite{affleck88,marston89}. While a negative $\kappa$ appeared to increase the stability of the dimerized phase, we found that this ground state is eventually destabilized with increasing $\kappa>0$ [see Fig.~\ref{fig:fmf+biq}(a)]. Above a critical $\kappa_c \approx 0.3$, the system enters a gapped chiral spin liquid with a $\pi/2$ flux on every triangular plaquette. 

We can gain a firmer grasp on the ground-state energetics within the pure $f$ sector and the result in Fig.~\ref{fig:fmf+biq}(a) by considering a few different mean-field ans\"atze. We start by setting $J_H = 1$ and diagonalizing
\begin{equation}
	\HH_\mathrm{MF}^f = E_0^f + \sum_{\vec{k}\sigma}
	\begin{pmatrix}
		f_{\vec{k}A\sigma}^\dagger & f_{\vec{k}B\sigma}^\dagger
	\end{pmatrix}
	\begin{pmatrix}
		\lambda_A - 2 \Re \left( e^{\ii k_1} \tilde{\chi}_{A1} \right) & - \sum_{n=2}^3 \left( e^{\ii k_n} \tilde{\chi}_{An} + e^{-\ii k_n} \tilde{\chi}_{Bn}^* \right)
		\\
		-\sum_{n=2}^3 \left( e^{\ii k_n} \tilde{\chi}_{Bn} + e^{-\ii k_n} \tilde{\chi}_{An}^* \right) & \lambda_B - 2 \Re \left( e^{\ii k_1} \tilde{\chi}_{B1} \right)
	\end{pmatrix}
	\begin{pmatrix}
		f_{\vec{k}A\sigma} \\
		f_{\vec{k}B\sigma}
	\end{pmatrix},
\end{equation}
from which we can obtain the spectrum
\begin{align}
	\epsilon_{\vec{k}\pm}^f &= \frac{(\lambda_A + \lambda_B)}{2} - \Re \left[ (\chit_{A1} + \chit_{B1}) e^{\ii k_1} \right]
	\notag \\
	&\pm \sqrt{
		\left\{ \frac{(\lambda_A - \lambda_B)}{2} - \Re \left[ (\chit_{A1} - \chit_{B1}) e^{\ii k_1} \right] \right\}^2 +  \abs{ \sum_{n=2}^3 \left( e^{\ii k_n} \chit_{An} + e^{-\ii k_n} \chit_{Bn}^* \right) }^2
	} \;.
	\label{eq:ekf general}
\end{align}
and the energy per site and flavor
\begin{align}
	\varepsilon_0 &= \frac{1}{N N_s} \lim_{T\to 0} \left[ E_0^f - TN \sum_{\vec{k}s} \ln( 1 + e^{- \epsilon_{\vec{k}s}^f/T } ) \right]
	\notag \\
	&=
	\frac{1}{2} \sum_a \left[ \sum_n \abs{\chi_{an}}^2 \left( 1 - 3\kappa \abs{\chi_{an}}^2 \right) - \frac{\lambda_a}{2} \right]
	+
	\frac{1}{N_s} \sum_{\vec{k}s} \epsilon_{\vec{k}s}^f \Theta\left( -\epsilon_{\vec{k}s}^f \right).
	\label{eq:eps0}
\end{align}

\subsubsection{Dimerized phase}

From the numerical solution of the mean-field equations, we find that a typical configuration for the dimerized (VBS) phase near $\kappa=0$ is characterized by parameters
\begin{align}
	\lambda_A &= \lambda_B = 0,
	&
	\chi_{A1} &= -\chi_{B1} = \chi_1 e^{\ii \phi_1},
	\notag \\
	\chi_{A2} &= \chi_2 e^{\ii \phi_2},
	&
	\chi_{B2} &= \chi_{A3} = \chi_{B3} = 0,
	\label{eq:ansatz dimerized}
\end{align}
with $0 < \chi_1 \ll \chi_2$. Substituting the ansatz \eqref{eq:ansatz dimerized} into Eq.~\eqref{eq:ekf general}, we find
\begin{equation}
	\epsilon_{\vec{k}\pm}^f = \pm \sqrt{ 4\chit_1^2 \cos^2 (k_1 + \phi_1) + \chit_2^2}.
\end{equation}
This shows that the bands are essentially flat in the limit $\chi_1/\chi_2 \to 0$, as one would expect for a strongly dimerized phase. The gap between the lower and upper flat bands is then given by $\Delta \approx J_H$.

\begin{figure}
	\centering
	\includegraphics[width=0.95\columnwidth]{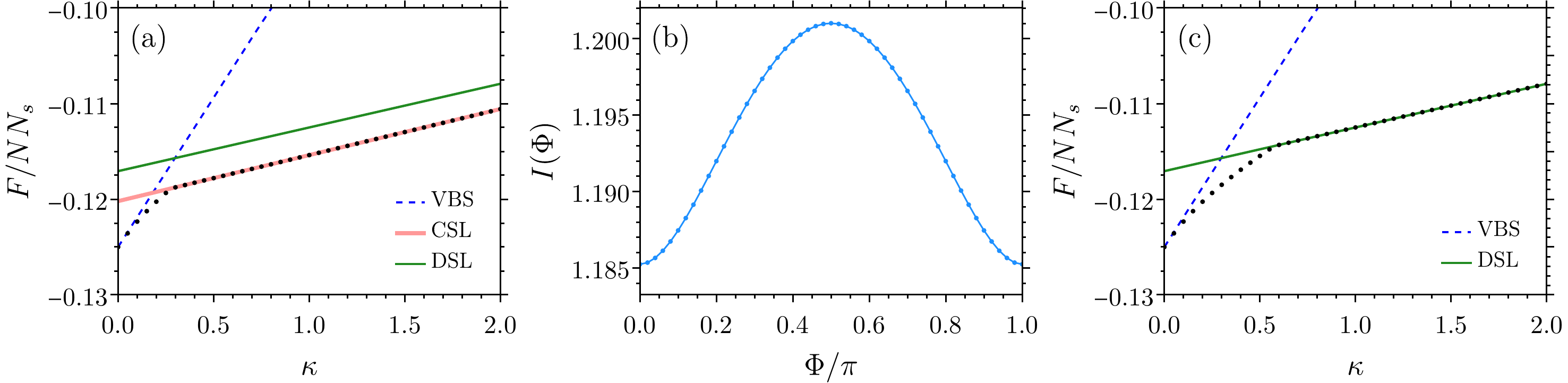}
	\caption{Mean-field results for the pure $f$ sector assuming a two-site unit cell. (a) Free energy per site and flavor as a function of the biquadratic coupling $\kappa$ for unrestricted, complex bond variables $\chi$. The black dots represent numerical data, whereas the colored lines correspond to analytic expressions obtained from different ans\"atze: a perfectly dimerized phase (VBS), a chiral spin liquid (CSL), and a Dirac spin liquid (DSL). (b) Function $I(\Phi)$ defined in Eq.~\eqref{eq:I}.The presence of a maximum at $\Phi = \pi/2$ shows that a chiral spin liquid is the most favorable among the variational states considered in Sec.~\ref{subsec:staggered}. (c) Same as (a), but for a restricted mean-field theory with real $\chi$ variables. This excludes the CSL solution and stabilizes a DSL for $\kappa \gtrsim 0.7$.}
	\label{fig:fmf+biq}
\end{figure}

Let us now compute the energy per site and flavor for a perfectly dimerized state, with $\chi_1=0$ and $\chi_2 = \chi$. Using Eq.~\eqref{eq:eps0}, we find that
\begin{align}
	\varepsilon_0 (\kappa, \chi)
	= \frac{1}{2} \left[ \chi^2(1 - 3\kappa \chi^2) - \abs{ \chi (1 - 2\kappa \chi^2) } \right].
\end{align}
Although $e_0$ is unbounded from below when $\kappa>0$, it has a stable local minimum at finite $\chi$ for each value of $\kappa$. One can show that this minimum occurs at $\chi = \pm 1/2$ for $\kappa < 2/3$ and $\chi = \pm 1/\sqrt{6\kappa}$ for $\kappa > 2/3$. Hence, the energy per site of a perfectly dimerized VBS state is
\begin{equation}
	\varepsilon_\mathrm{VBS} \left(\kappa\right) =
	\begin{dcases}
		\frac{1}{8} \left( \frac{\kappa}{4} - 1 \right) & \text{for } \kappa \le 2/3,
		\\
		\frac{1}{6} \left( \frac{1}{4\kappa} - \sqrt{ \frac{2}{3\kappa} } \right) & \text{for } \kappa > 2/3,
	\end{dcases}
	\label{eq:evbs}
\end{equation}
which is plotted as a dashed blue line in Fig.~\ref{fig:fmf+biq}(a). Note that Eq.~\eqref{eq:evbs} yields the correct result $\varepsilon_0 = -0.125$ for $\kappa = 0$, but deviates from the numerical solutions for $\kappa > 0$ due to the oversimplified assumption that $\chi_1 = 0$. Indeed, our numerical results show that the dimerization becomes imperfect for any $\kappa > 0$.

\subsubsection{Staggered $\Phi$ and $\pi - \Phi$ fluxes}
\label{subsec:staggered}

Next, we consider the class of ans\"atze
\begin{align}
	\lambda_A &= \lambda_B = 0,
	&
	\chi_{A1} &= \chi_{A3} = \chi
	\notag
	\\
	\chi_{A2} &= \chi_{B2} = \chi e^{\ii\Phi},
	&
	\chi_{B1} &= \chi_{B3} = -\chi,
	\label{eq:staggered ansatz}
\end{align}
with $\chi, \Phi \in \mathbb{R}$. Using Eq.~\eqref{eq:fluxes}, one finds that this yields a staggered flux configuration with $\Phi_1 = \Phi_3 = \Phi$ and $\Phi_2 = \Phi_4 = \pi - \Phi$. The dispersion and energy per site and flavor are given by
\begin{align}
	\epsilon_{\vec{k}\pm}^f &= \pm 2 \abs{\chi (1-2\kappa\chi^2)} \sqrt{\cos^2 k_1 + \cos^2 (k_2 + \Phi) + \sin^2 k_3},
	\label{eq:staggered dispersion}
	\\
	\varepsilon_0 (\kappa, \chi, \Phi) &= 3\chi^2 (1-3\kappa \chi^2) - I(\Phi) \abs{\chi (1-2\kappa\chi^2)},
	\label{eq:staggered energy}
\end{align}
where the function
\begin{align}
	I(\Phi) &= \frac{1}{N_c} \sum_\vec{k} \sqrt{ \cos^2 k_1 + \cos^2 (k_2 + \Phi) + \sin^2 k_3 }.
	\label{eq:I}
\end{align}
is plotted in Fig.~\ref{fig:fmf+biq}(b). With this, one can verify that $\Phi = 0$ gives rise to a Dirac spin liquid (DSL) with nodes at $\vec{K}_\pm = \pm ( \frac{\pi}{2} , \frac{\pi}{2\sqrt{3}} )$, whereas $\Phi = \pi/2$ corresponds to a chiral spin liquid (CSL) with a gapped spectrum and uniform fluxes $\Phi_n = \pi/2$.

It is straightforward to show that, while Eq.~\eqref{eq:staggered energy} is unbounded from below, it has a local minimum at $\chi = \pm I(\Phi)/6$ for $\kappa < 6/I^2(\Phi)$ and $\chi = 1/\sqrt{6\kappa}$ for $\kappa > 6/I^2(\Phi)$. The energy of the staggered-flux solution is thus given by
\begin{align}
	\varepsilon_\mathrm{stagg} \left( \Phi, \kappa \right) &= 
	\begin{dcases}
		-\frac{I^2 (\Phi)}{12} \left( 1 - \frac{\kappa}{36} I^2(\Phi) \right) & \text{for } \kappa \le 6/I^2,
		\\
		\frac{1}{4\kappa} - \frac{I(\Phi)}{9} \sqrt{ \frac{6}{\kappa} } & \text{for } \kappa > 6/I^2.
	\end{dcases}
	\label{eq:stagg energy}
\end{align}
From this, it becomes clear that the most competitive state in this class of ans\"atze must be the one to \emph{maximize} the function $I(\Phi)$. This is exactly the CSL solution mentioned above, whose energy $\varepsilon_\mathrm{stagg} (\pi/2, \kappa)$ is plotted as a red curve in Fig.~\ref{fig:fmf+biq}(a). For comparison, we also show the energy $\varepsilon_\mathrm{stagg} (0, \kappa)$ of the DSL in green.

\subsection{Pure $f$ sector: Restricted mean-field theory}

The previous subsection showed that a DSL does not naturally emerge as a ground state of our full large-$N$ mean field theory. We found that this remains the case even after including a next-nearest-neighbor Heisenberg interaction, although recent numerical studies indicate that this is sufficient to stabilize a DSL in an SU(2)-symmetric system \cite{iqbal16,hu19,sherman23,wietek24}.

To obtain this state as a self-consistent solution, we can restrict ourselves to real $\chi$ variables, and thus eliminate the CSL solution. For small $\kappa$, we still obtained a dimerized phase; however, as shown in Fig.~\ref{fig:fmf+biq}(c), the DSL described above becomes the ground state for large $\kappa \gtrsim 0.7$. The spectrum of this state and its energy per site and flavor are given exactly by Eqs.~\eqref{eq:staggered dispersion} and \eqref{eq:stagg energy} with $\Phi = 0$.

\subsection{Pure $c$ sector: Evolution of the bare electronic Fermi surface}

\begin{figure}
	\centering
	\includegraphics[width=\columnwidth]{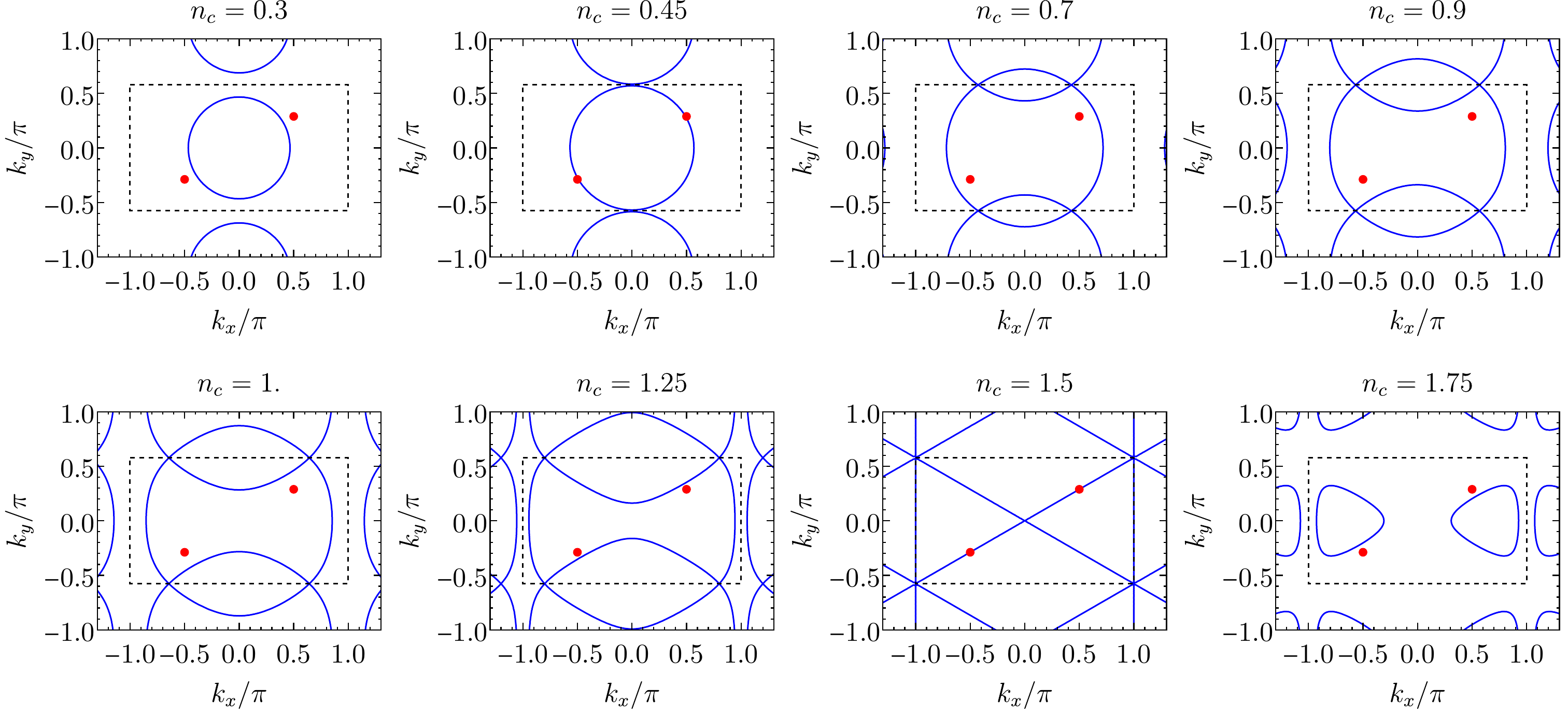}
	\caption{Bare $c$ Fermi surfaces for different fillings. The Brillouin zone associated with the two-site mean-field unit cell is shown in dashed lines, and the momenta $\vec{K}_\pm = \pm (\frac{\pi}{2}, \frac{\pi}{2\sqrt{3}})$ corresponding to the locations of the Dirac cones of the Dirac spin liquid are highlighted in red circles.}
	\label{fig:cfs}
\end{figure}

To characterize the evolution of the bare $c$-electron Fermi surface as a function of the filling $n_c$, we diagonalize the hopping Hamiltonian in Eq.~\eqref{eq:Ht}. This yields the bare electronic spectrum $\epsilon_{\vec{k}\pm}^c = -2t \left[ \cos k_1 \pm \abs{\cos k_2 + \cos k_3} \right]$. Using the condition $n_c = \frac{1}{N} \sum_{ia\sigma} \expval{c_{ia\sigma}^\dagger c_{ia\sigma}}$, we then solve for the chemical potential $\mu$ for various fillings $n_c$, thereby obtaining the Fermi surfaces shown in Fig.~\ref{fig:cfs}. Note that the Fermi surfaces intersect the nodes of the DSL described above at $n_c \approx 0.45$ and $n_c = 1.5$.

\begin{figure}[th!]
	\centering
	\includegraphics[width=\columnwidth]{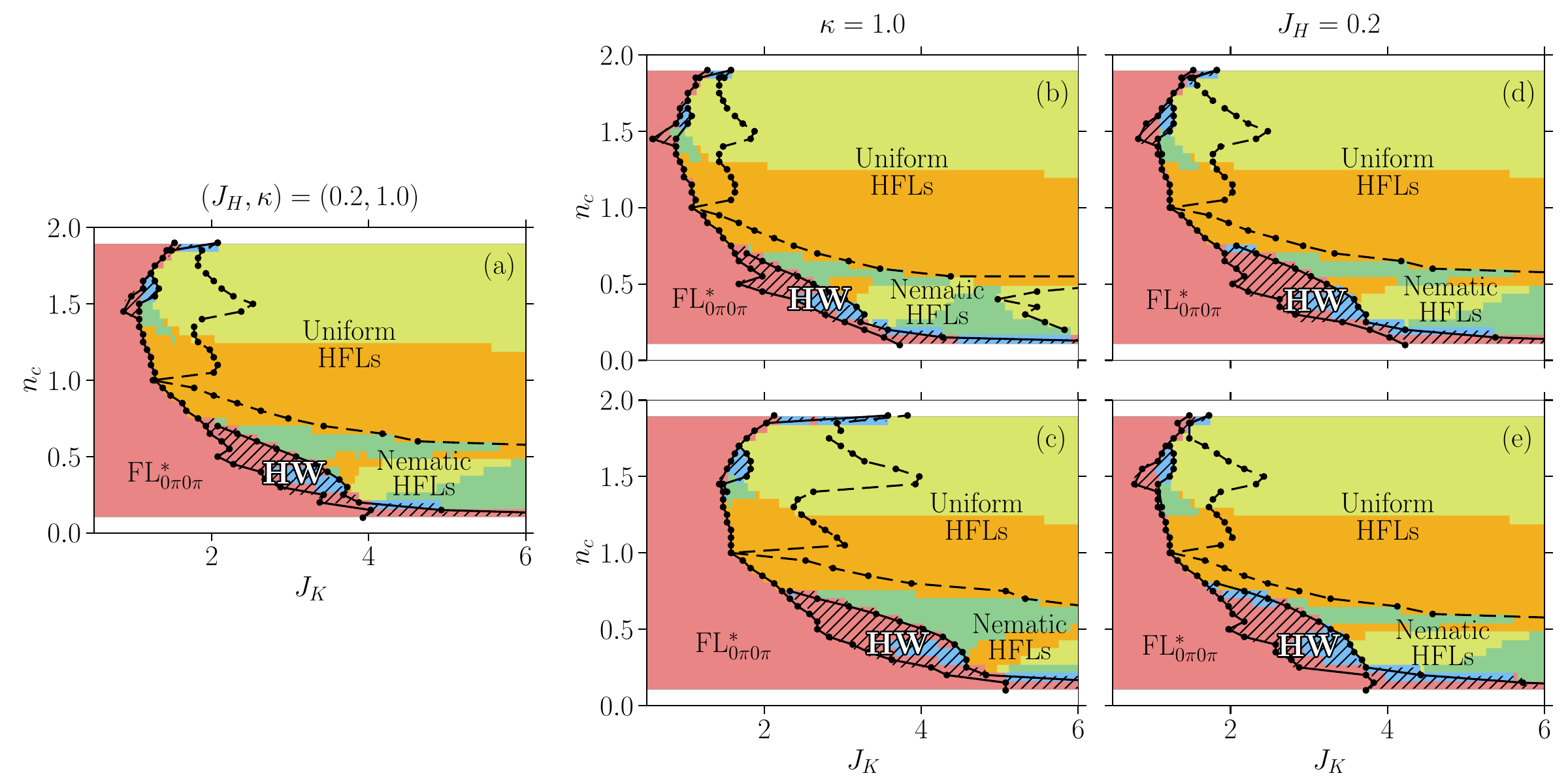}
	\caption{Mean-field phase diagrams for $L=50$, $k_\mathrm{B} T = 10^{-3}$, and varying values of $(J_H, \kappa)$. Panel (a) uses the same parameters $(J_H, \kappa) = (0.2, 1.0)$ as Fig.~2(b) in the main text, although at a smaller system size. Panels (b,c) have fixed $\kappa = 1.0$ with (b) $J_H = 0.1$ and (c) $J_H = 0.5$. Conversely, panels (d,e) have fixed $J_H = 0.2$ with (d) $\kappa = 2.0$ and (e) $\kappa = 3.0$. The colors, lines, and hatching in the plots have the same meaning as those in Fig.~2(b) in the main text.}
	\label{fig:pds}
\end{figure}

\begin{figure}[h!]
	\centering
	\includegraphics[width=\columnwidth]{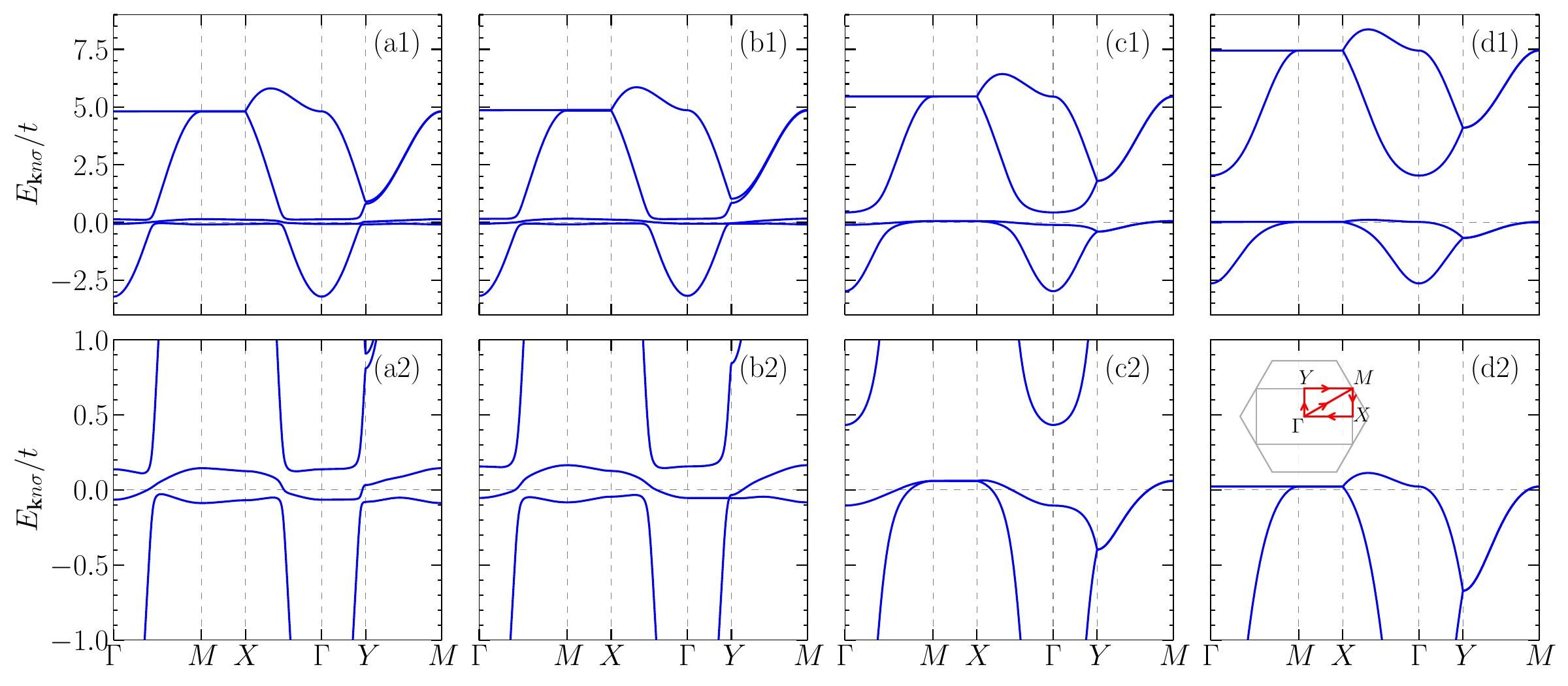}
	\caption{Mean-field band structure for four different states obtained at fixed $L=150$, $k_B T = 10^{-3}$, $J_H = 0.2$, $\kappa = 1$, and $n_c = 0.35$: (a) HW$_{0\pi 0\pi}$, (b) HW$_{0\pi\pi 0}$, (c) a zero-flux heavy Fermi liquid with nematic $\abs{\chi}$, and (d) a zero-flux heavy Fermi liquid with uniform $\abs{\chi}$. These states were obtained as self-consistent solutions of the mean-field theory at $J_K = 2.95$, $3.00$, $4.50$, and $8.00$, respectively. The bottom row shows the same data as the top row, but zoomed in around the Fermi level. The inset in (d2) shows the momentum-space path used for the plots. There, the gray hexagon and rectangle respectively represent the first Brillouin zone of the triangular lattice and a reduced Brillouin zone associated with the two-site unit cell.}
	\label{fig:mfbands}
\end{figure}

\subsection{Further mean-field results}

In this subsection, we present additional mean-field results, reflecting (i) how variations in the Heisenberg coupling $J_H$ and the biquadratic interaction $\kappa$ affect the phase diagram and (ii) the difference between the band structures of self-consistent HW and heavy-Fermi-liquid states.

Representative phase diagrams obtained for $L=50$ and $k_B T = 10^{-3}$ are shown in Fig.~\ref{fig:pds}, where (a) corresponds to the parameter set utilized in Fig.~2(b) of the main text, $(J_H, \kappa) = (0.2, 1.0)$. 
The middle column in Fig.~\ref{fig:pds} illustrates the effect of varying $J_H$ at fixed $\kappa = 1$ by showing data for (b) $J_H = 0.1$ and (c) $J_H = 0.5$.  As expected, larger values of $J_H$ lead to a more stable FL$^*$, with higher critical couplings $J_K^c (n_c)$. Aside from that, we also verify an expansion in the domains of the nematic states with uniform $b_a$ and the HWs from (b) $\to$ (a) $\to$ (c).
Conversely, the last column in Fig.~\ref{fig:pds} depicts the effect of changing the biquadratic interaction to (d) $\kappa = 2$ and (e) $\kappa = 3$ at fixed $J_H = 0.2$. This time, we do not observe any salient features, except for a slight growth of the HW$_{0\pi\pi0}$ phase (light blue) in Fig.~\ref{fig:pds}(e).

Meanwhile, Fig.~\ref{fig:mfbands} depicts the band structure of self-consistent mean-field solutions obtained at fixed $L=150$, $k_B T = 10^{-3}$, $J_H = 0.2$, $\kappa = 1$, and $n_c = 0.35$. The results are organized in order of increasing $J_K$ from left to right. This explains the increase of the gap between the conduction and valence bands from (a) to (d). Furthermore, the ``flat bands'' of the HW states in (a) and (b) are noticeably more dispersive than their counterparts in the heavy-Fermi-liquid states with (c) nematic and (d) uniform values of $\abs{\chi}$ because the kinetic energy of the spinons is quenched as $J_K/J_H$ grows and the system distances itself from the FL$^*_{0\pi0\pi}$ phase.


\section{Further details on the perturbative analysis}
\label{sec:pt}

In this section, we expand on the End Matter by providing further details on our perturbative analysis, which is geared to show how Kondo screening emerges as the system exits the FL$^*$ phase via a continuous quantum phase transition.
After outlining the derivation of the polarization bubble $\Pi (\vec{q}, \ii\nu_p)$ (Sec.~\ref{subsec:Pider}), we provide a short proof that the leading Gaussian instability out of the FL$^*$ phase occurs for $\nu_p = 0$ (Sec.~\ref{subsec:proof}).
Subsequently, we identify two different projective symmetries of the Kondo-Heisenberg action $S$ written in Eq.~(6) of the main text (Sec.~\ref{subsec:symmetries}) and show how these symmetries constrain the form of Landau free energy that describes the onset of screening (Sec.~\ref{subsec:landau}).

\subsection{Derivation of the polarization bubble}
\label{subsec:Pider}

The first step required to compute $\Pi (\vec{q}, \ii\nu_p)$ is to derive the free propagators $G_{\ell 0}^{aa'} (\vec{k}, \ii\omega_n) = -\expval{\ell_{\vec{k}a\sigma} (\ii\omega_n) \ell^*_{\vec{k}a'\sigma} (\ii\omega_n)}_0$, where $\ell = c,f$ and $\expval{\cdots}_0$ denotes the thermodynamic average with respect to the noninteracting action $S_0$. 
To this end, we diagonalize the Hermitian matrices
\begin{equation}
	h_{\ell\vec{k}} = 
	\begin{pmatrix}
		h_{\ell\vec{k}}^{11} & h_{\ell\vec{k}}^{12}
		\\
		\left( h_{\ell\vec{k}}^{12} \right)^* & h_{\ell\vec{k}}^{22}
	\end{pmatrix}
	\label{eq:hlk}
\end{equation}
describing the noninteracting $c$ and $f$ sectors so as to obtain their eigenvalues and orthonormal eigenvectors:
\begin{align}
	\xi_{\vec{k}}^{\ell \pm} &=
	\frac{h_{\ell\vec{k}}^{11} + h_{\ell\vec{k}}^{22}}{2}
	\pm
	\sqrt{
	\left( \frac{h_{\ell\vec{k}}^{11} - h_{\ell\vec{k}}^{22}}{2} \right)^2 + \abs{h_{\ell\vec{k}}^{12}}^2
	},
	&
	\ket{u_{\ell \vec{k} \pm}} &=
	\frac{1}{\sqrt{
		\left( \xi_{\vec{k}}^{\ell \pm} - h_{\ell\vec{k}}^{11} \right)^2 + \abs{h_{\ell\vec{k}}^{12}}^2
	}}
	\begin{pmatrix}
		h_{\ell\vec{k}}^{12}
		\\
		\xi_\vec{k}^{\ell\pm} - h_{\ell\vec{k}}^{11}
	\end{pmatrix}.
\end{align}
Then, denoting the $a$-th component of $\ket{u_{\ell\vec{k}s}}$ by $u_{\ell \vec{k} s}^{(a)}$, we obtain
\begin{align}
	G_{\ell 0}^{aa'} (\vec{k}, \ii\omega_n) &= 
	\sum_{s} \frac{ u_{\ell\vec{k} s}^{(a)} u_{\ell\vec{k} s}^{(a') *} }{\ii\omega_n - \xi_{\vec{k}}^{\ell s}}.
	\label{eq:Gl0}
\end{align}

Substituting Eq.~\eqref{eq:Gl0} into the definition of $\Pi$ yields
\begin{align}
	\Pi (\vec{q}, \ii\nu_p) &= 
	-\frac{1}{\beta N_c} \sum_{\vec{k}} \sum_{\ii\omega_n} \sum_{ss'}
	\frac{\ket{w_{ss'} (\vec{k}, \vec{q}) } \bra{w_{ss'} (\vec{k}, \vec{q})}}{(\ii\omega_n + \ii\nu_p - \xi_{\vec{k}+\vec{q}}^{cs})(\ii\omega_n - \xi_\vec{k}^{fs'})}
	\label{eq:Pimat1},
\end{align}
with form factors $w_{ss'}^{(a)} (\vec{k}, \vec{q}) = u_{c, \vec{k}+\vec{q}, s}^{(a)} u_{f, \vec{k}, s'}^{(a)*}$. Using standard techniques of complex analysis, one can evaluate the sum over Matsubara frequencies and arrive at
\begin{align}
	\Pi (\vec{q}, \ii\nu_p) &= 
	\frac{1}{N_c} \sum_{\vec{k}} \sum_{ss'}
	\frac{
		n_F \left( \xi_{\vec{k}}^{fs'} \right) - n_F \left( \xi_{\vec{k}+\vec{q}}^{cs} \right)
	}{
		\xi_{\vec{k}+\vec{q}}^{cs} - \xi_{\vec{k}}^{fs'} - \ii\nu_p
	}
	\ket{w_{ss'} (\vec{k}, \vec{q}) } \bra{w_{ss'} (\vec{k}, \vec{q})},
	\label{eq:Pimat2}
\end{align}
where $n_F$ denotes the Fermi-Dirac distribution function. We thus see that the polarization operator is given by a weighted sum of Lindhard functions that describe particle-hole transitions between the $c$ and $f$ sectors. From this, it is easy to isolate the Hermitian and anti-Hermitian components of the polarization matrix:
\begin{align}
	\Pi_H (\vec{q},\ii\nu_p)&= \frac{\Pi + \Pi^\dagger}{2} 
	= 
	\frac{1}{N_c} \sum_{\vec{k}} \sum_{ss'}
	\frac{
		\left[ n_F \left( \xi_{\vec{k}}^{fs'} \right) - n_F \left( \xi_{\vec{k}+\vec{q}}^{cs} \right) \right]
		( \xi_{\vec{k}+\vec{q}}^{cs} - \xi_{\vec{k}}^{fs'} )
	}{
		( \xi_{\vec{k}+\vec{q}}^{cs} - \xi_{\vec{k}}^{fs'} )^2 + \nu_p^2
	}
	\ket{w_{ss'} (\vec{k}, \vec{q}) } \bra{w_{ss'} (\vec{k}, \vec{q})},
	\label{eq:PiHmat}
	\\
	\Pi_A (\vec{q},\ii\nu_p)&= \frac{\Pi - \Pi^\dagger}{2} 
	= 
	\frac{\ii\nu_p}{N_c} \sum_{\vec{k}} \sum_{ss'}
	\frac{
		n_F \left( \xi_{\vec{k}}^{fs'} \right) - n_F \left( \xi_{\vec{k}+\vec{q}}^{cs} \right)
	}{
		( \xi_{\vec{k}+\vec{q}}^{cs} - \xi_{\vec{k}}^{fs'} )^2 + \nu_p^2
	}
	\ket{w_{ss'} (\vec{k}, \vec{q}) } \bra{w_{ss'} (\vec{k}, \vec{q})}.
\end{align}
The last equation shows that $\Pi$ is Hermitian at zero frequency, but generally has a nonzero anti-Hermitian part.

\subsubsection{$T\to 0$ limit}

Despite its compact form, the expression for the polarization operator in Eq.~\eqref{eq:Pimat2} is not very transparent, because it encodes substantial information in the form factors $\ket{w_{ss'}}\bra{w_{ss'}}$. We now show how, using basic symmetries of $h_{c\vec{k}}$ and $h_{f\vec{k}}$, one can unravel such hidden information and derive relatively simple results for $\Pi(\vec{q}) \equiv \Pi(\vec{q},0)$ in the limit $T\to 0$.

Since the $c$-electron hopping Hamiltonian is assumed not to break the $A \leftrightarrow B$ sublattice symmetry, the matrix elements in Eq.~\eqref{eq:hlk} cannot all be independent. We must demand that $h_{c\vec{k}}^{22} = h_{c\vec{k}}^{11}$, and hence
\begin{align}
	\xi_{\vec{k}}^{c\pm} &= h_{c\vec{k}}^{11} \pm \abs{h_{c\vec{k}}^{12}},
	&
	\ket{u_{c\vec{k}s}} &= \frac{1}{\sqrt{2}}
	\begin{pmatrix}
		h_{c\vec{k}}^{12} / \abs{h_{c\vec{k}}^{12}}
		\\
		s
	\end{pmatrix}.
	\label{eq:ceig}
\end{align}
On the other hand, if we require the bare $f$ spectrum to be particle-hole symmetric ($\xi_{\vec{k}}^{f\pm} = \pm \eps_\vec{k}^f$) as in the DSL ansatz, then the noninteracting $f$ Bloch Hamiltonian must be of the form $h_{f\vec{k}} = \vec{d}_\vec{k} \cdot \bm{\sigma}$, where $\bm{\sigma} = (\sigma_x, \sigma_y, \sigma_z)$ is a vector of Pauli matrices. This implies that $h_{f\vec{k}}^{22} = -h_{f\vec{k}}^{11}$ and
\begin{align}
	\xi_{\vec{k}}^{f\pm} &= \pm \eps_{\vec{k}}^f = 
	\pm \sqrt{ \abs{h_{f\vec{k}}^{11}}^2 + \abs{h_{f\vec{k}}^{12}}^2 },
	&
	\ket{u_{f\vec{k}s}} &= \frac{1}{ \sqrt{2\eps_{\vec{k}}^f \left( \eps_{\vec{k}}^f - s h_{f\vec{k}}^{11} \right)} }
	\begin{pmatrix}
		h_{f\vec{k}}^{12} 
		\\
		s \left( \eps_{\vec{k}}^f - s h_{f\vec{k}}^{11} \right)
	\end{pmatrix}.
	\label{eq:feig}
\end{align}

Taking the limit $T\to 0$ with $\xi_{\vec{k}}^{f\pm} = \pm \eps_\vec{k}^f$, we can simplify Eq.~\eqref{eq:Pimat2} by setting $n_F(-\eps_\vec{k}^f) = 1$ and $n_F(\eps_\vec{k}^f) = 0$. The result reads
\begin{align}
	\Pi_{aa'} (\vec{q}) &= \frac{1}{N_c} \sum_{\vec{k}s} 
	\frac{
		u_{c,\vec{k}+\vec{q},s}^{(a)} u_{c,\vec{k}+\vec{q},s}^{(a')*}
	}{
		\eps_\vec{k}^f + \abs{\xi_{\vec{k}+\vec{q}}^{cs}}
	}
	\left[ 
	u_{f,\vec{k},-}^{(a)^*} u_{f,\vec{k},-}^{(a')} \Theta\left( \xi_{\vec{k} + \vec{q}}^{cs} \right) 
	+
	u_{f,\vec{k},+}^{(a)^*} u_{f,\vec{k},+}^{(a')} \Theta\left( -\xi_{\vec{k} + \vec{q}}^{cs} \right) 
	\right],
\end{align}
where $\Theta$ is the Heaviside function. Note that the first (second) term in the square brackets describes virtual transitions from occupied states of an $f$ ($c$) band to unoccupied states in a $c$ ($f$) band. Using Eqs.~\eqref{eq:ceig} and \eqref{eq:feig}, one can show that the three contributions entering the function $p_+ (\vec{q})$ in Eq.~(5) of the main text are given by Eqs.~(11a)--(11c) in the main text.

\subsection{Why the Gaussian instability occurs at $\nu_p= 0$}
\label{subsec:proof}

The results derived in the previous subsection evince that the quadratic kernel
\begin{equation}
	R(J_K, \vec{q}, \ii\nu_p) = NJ_K^2 \left[ \frac{1}{J_K} \mathds{1} - \Pi(\vec{q}, \ii\nu_p) \right]
	\label{eq:kernelK}
\end{equation}
of the effective action $S_b$ is generally non-Hermitian. However, the stability of the Gaussian theory is conditioned \emph{exclusively} by the Hermitian part $R_H = (R + R^\dagger)/2$. Specifically, the functional integral $Z \sim \int \mathcal{D} (b^*, b) e^{-S_b}$ converges as long as $R_H(J_K, q)$ is positive definite for every $q$ \cite{negele_orland}. The violation of this condition indicates the condensation of $b$ and the need to include higher-order terms to construct a stable action.

Suppose that, as $J_K$ increases, $R_H$ first ceases to be positive definite at a coupling $J_K = J_K'$ and momentum $Q = (\vec{Q}, \ii\nu_Q)$, with $\nu_Q \ne 0$. Concretely, this implies that $R_H$ acquires a zero mode $\ket{v_0}$, and hence
\begin{equation}
	\expval{R_H (J_K', \vec{Q}, \ii\nu_Q)}{v_0} = 
	NJ_K'^2 \left[ \frac{1}{J_K'} - \expval{\Pi_H (\vec{Q}, \ii\nu_Q)}{v_0} \right] = 0.
	\label{eq:proof1}
\end{equation}
However, one can use Eq.~\eqref{eq:PiHmat} to show that
\begin{align}
	\expval{\Pi(\vec{q},0) - \Pi_H (\vec{q}, \ii\nu_p)}{v} &=
	\frac{1}{N_c} \sum_{\vec{k}} \sum_{ss'}
	\left[
	\frac{n_F( \xi_{\vec{k}}^{fs'}) - n_F(\xi_{\vec{k}+\vec{q}}^{cs})}{ \xi_{\vec{k}+\vec{q}}^{cs} - \xi_{\vec{k}}^{fs'} }
	\right]
	\frac{\nu_p^2}{ ( \xi_{\vec{k}+\vec{q}}^{cs} - \xi_{\vec{k}}^{fs'} )^2 + \nu_p^2}
	\abs{ \braket{w_{ss'} (\vec{k}, \vec{q})}{v} }^2
	\notag \\
	&\geq 0
	\label{eq:proof2}
\end{align}
for an arbitrary vector $\ket{v}$. Note that the inequality follows from the fact that the function in the square brackets is positive and that the equality only holds if $\nu_p = 0$. Since we assumed that $\nu_Q \ne 0$, Eqs.~\eqref{eq:proof1} and \eqref{eq:proof2} lead to a contradiction:
\begin{align}
	\expval{R_H (J_K', \vec{Q}, 0)}{v_0} &= 
	NJ_K'^2 \left[
	\frac{1}{J_K'} - \expval{\Pi (\vec{Q}, 0)}{v_0}
	\right]
	\notag \\
	&< 
	NJ_K'^2 \left[ \frac{1}{J_K'} - \expval{\Pi_H (\vec{Q}, \ii\nu_Q)}{v_0} \right] = 0,
\end{align}
Thus, regardless of the value of $\nu_Q \ne 0$, $R_H(J_K, \vec{Q}, 0)$ always ceases to be positive definite at a coupling $J_K^c < J_K'$. This proves that the true instability must occur at zero frequency, a conclusion that agrees with the notion that an equilibrium phase transition is signaled by the onset of a static observables.

\subsection{Symmetries of the Kondo-Heisenberg action}
\label{subsec:symmetries}

Let us now turn to a symmetry analysis of the full Kondo-Heisenberg action in Eq.~(6) of the main text. We shall start by considering the effect of time reversal $\mathcal{T}$ in the physical $N=2$ case, where $\mathcal{T}^2 = -1$. Similar results nonetheless follow for $N>2$ generalizations if one assumes spinless time reversal, $\mathcal{T}^2 = +1$. The action of $\mathcal{T}$ on fields in $S$ is given by
\begin{align}
	\mathcal{T}: \quad
	c_{ia\sigma} (\tau) &\mapsto \sum_{\sigma'} (\ii \sigma_y)_{\sigma\sigma'} c_{ia\sigma'} (\tau),
	&
	c_{\vec{k} a\sigma} (\tau) &\mapsto \sum_{\sigma'} (\ii \sigma_y)_{\sigma\sigma'} c_{-\vec{k}, a\sigma'} (\tau),
	\notag \\
	f_{ia\sigma} (\tau) &\mapsto \sum_{\sigma'} (\ii \sigma_y)_{\sigma\sigma'} f_{ia\sigma'} (\tau),
	&
	f_{\vec{k} a\sigma} (\tau) &\mapsto \sum_{\sigma'} (\ii \sigma_y)_{\sigma\sigma'} f_{-\vec{k}, a\sigma'} (\tau),
	\notag \\
	b_{ia} (\tau) &\mapsto b_{ia} (\tau),
	&
	b_{\vec{q} a} (\tau) &\mapsto b_{-\vec{q}, a} (\tau).
\end{align}
One can thus verify that the Kondo term is invariant under $\mathcal{T}$ and, thanks to the property $h_{c,-\vec{k}} = h_{c\vec{k}}$, which is indeed satisfied by Eq.~(7), so is the quadratic term in the $c$ fermions. On the other hand, the quadratic part in the $f$ fermions transforms as:
\begin{align}
	\sum_{\vec{k}\sigma} f_{\vec{k} a\sigma}^* h_{f\vec{k}}^{aa'} f_{\vec{k} a'\sigma} 
	&\longmapsto
	\sum_{\vec{k}\sigma} f_{\vec{k} a\sigma}^* [h_{f,-\vec{k}}^{aa'}]^* f_{\vec{k} a'\sigma}.
	\label{eq:fquadTheta}
\end{align}
From Eq.~(8), one finds that $h_{f,-\vec{k},} = h_{f\vec{k}}^*$. Hence, Eq.~\eqref{eq:fquadTheta} and the entire action $S$ are invariant under $\mathcal{T}$.

Another symmetry of interest is related to a translation $T_{\vec{a}_2/2}$ by half of the primitive vector $\vec{a}_2 = (1,\sqrt{3})$, which exchanges the sublattices $A \leftrightarrow B$ introduced by the DSL ansatz [see Fig.~2(a)]. If $\vec{r}_{j\bar{a}} = \vec{r}_{ia} + \vec{a}_2/2$, then this transformation acts on $S$ according to
\begin{align}
	T_{\vec{a}_2/2} : \quad
	c_{ia\sigma} (\tau) &\mapsto c_{j\bar{a}\sigma} (\tau),
	&
	c_{\vec{k} a\sigma} (\tau) &\mapsto e^{\ii \vec{k}\cdot \vec{a}_2/2} c_{\vec{k} \bar{a} \sigma} (\tau),
	\notag \\
	f_{ia\sigma} (\tau) &\mapsto f_{j\bar{a}\sigma} (\tau),
	&
	f_{\vec{k} a\sigma} (\tau) &\mapsto e^{\ii \vec{k}\cdot \vec{a}_2/2} f_{\vec{k} \bar{a} \sigma} (\tau),
	\notag \\
	b_{ia} (\tau) &\mapsto b_{j\bar{a}} (\tau),
	&
	b_{\vec{q} a} (\tau) &\mapsto e^{\ii \vec{q}\cdot \vec{a}_2/2} b_{\vec{q} \bar{a}} (\tau).
	\label{eq:Ta2ov2}
\end{align}
Since the Kondo interaction is local and the $c$-electron hopping Hamiltonian is periodic in $\vec{a}_2/2$, the corresponding terms in $S$ are invariant under Eq.~\eqref{eq:Ta2ov2}. However, the quadratic part in $f$ transforms nontrivially:
\begin{align}
	T_{\vec{a}_2/2}: \quad
	\sum_{\vec{k}\sigma} f_{\vec{k} a\sigma}^* h_{f\vec{k}}^{aa'} f_{\vec{k} a'\sigma} 
	\longmapsto
	\sum_{\vec{k}\sigma} f_{\vec{k} a\sigma}^* h_{f\vec{k}}^{\bar{a}\bar{a}'} f_{\vec{k} a'\sigma} .
\end{align}
To compensate for the change in the matrix $h_{f\vec{k}}$, we can exploit the gauge redundancy introduced by the parton representation in Eq.~\eqref{eq:partonrep} and apply the local gauge transformations
\begin{align}
	U_{g} : \quad
	f_{ia\sigma} (\tau) &\mapsto e^{-\ii \frac{\vec{b}_1}{2} \cdot \vec{r}_{ia}} f_{ia\sigma} (\tau),
	&
	f_{\vec{k} a\sigma} (\tau) &\mapsto f_{\vec{k} + \vec{b}_1/2, a, \sigma} (\tau),
	\notag \\
	b_{ia} (\tau) &\mapsto e^{i \frac{\vec{b}_1}{2} \cdot \vec{r}_{ia}} b_{ia} (\tau),
	&
	b_{\vec{q}a} (\tau) &\mapsto b_{\vec{q} - \vec{b}_1/2, a} (\tau),
	\label{eq:Ug}
\end{align}
where $\vec{b}_1 = (2\pi, -2\pi/\sqrt{3})$ is the reciprocal lattice vector that obeys $\vec{a}_i \cdot \vec{b}_1 = 2\pi \delta_{i1}$ for $i=1,2$. As a local gauge transformation, Eq.~\eqref{eq:Ug} preserves the flux configuration of the DSL. Furthermore, it leaves each contribution $b_{ia} c_{ia\sigma}^* f_{ia\sigma}$ in the Kondo term invariant while restoring the quadratic part in $f$ to its original form. Indeed, if $f_{\vec{k}\sigma} = \begin{pmatrix} f_{\vec{k}A\sigma} & f_{\vec{k}B\sigma} \end{pmatrix}^\top$, then
\begin{align}
	U_{g} T_{\vec{a}_2/2}: \quad
	\sum_{\vec{k}\sigma} f_{\vec{k} \sigma}^* h_{f\vec{k}} f_{\vec{k} \sigma}
	&\longmapsto
	\sum_{\vec{k}\sigma} f_{\vec{k} \sigma}^*
	\begin{pmatrix}
		h_{f, \vec{k} - \vec{b}_1/2}^{22} & h_{f,\vec{k} - \vec{b}_1/2}^{21}
		\\
		h_{f, \vec{k} - \vec{b}_1/2,}^{12} & h_{f,\vec{k} - \vec{b}_1/2}^{11}
	\end{pmatrix} 
	f_{\vec{k} \sigma} 
	\notag \\
	&\longmapsto
	\sum_{\vec{k}\sigma} f_{\vec{k} \sigma}^* h_{f\vec{k}} f_{\vec{k} \sigma}.
	\label{eq:UgT}
\end{align}
Here, the last step follows from using Eq.~(8) in the main text (i.e., a property of the DSL ansatz) together with the ensuing properties $h_{f, \vec{k} \pm \vec{b}_1/2}^{22} = -h_{f\vec{k}}^{22} = h_{f\vec{k}}^{11}$ and $h_{f, \vec{k} \pm \vec{b}_1/2}^{12} = (h_{f\vec{k}}^{12})^* = h_{f\vec{k}}^{21}$. Thus, we conclude that $U_{g} T_{\vec{a}_2/2}$ is a symmetry of $S$.

\subsection{Landau free energy for the hybridization fields}
\label{subsec:landau}

The fact that the full Kondo-Heisenberg action $S$ possesses a given symmetry implies that the same symmetry must be present at every order of the effective action $S_b = S_b^{(2)} + S_b^{(4)} + \cdots$. In particular, it must apply to
\begin{equation}
	S_b^{(2)} = \sum_{q}
	\begin{pmatrix}
		b_A^* (q) & b_B^* (q)
	\end{pmatrix}
	R(q)
	\begin{pmatrix}
		b_A (q) \\ b_B (q)
	\end{pmatrix}
	\equiv
	\sum_q
	b^\dagger (q) R(q) b(q),
\end{equation}
where the kernel $R(q)$ is given by Eq.~\eqref{eq:kernelK}. 

The invariance of $S$ under time reversal requires that
\begin{equation}
	S_b^{(2)} \overset{!}{=}
	\sum_q
	b^\dagger (-q) R^*(q) b(-q)
	=
	\sum_q
	b^\dagger (q) R^*(-q) b(q),
\end{equation}
or, equivalently, that $R^*(-q) = R(q)$. Since $R(\vec{q}) \equiv R(\vec{q},0)$ is Hermitian and obeys an eigenvalue equation $R(\vec{q}) v_\alpha(\vec{q}) = r_{\alpha}(\vec{q}) v_\alpha(\vec{q})$ with real $r_\alpha$ for every $\vec{q}$ and $\alpha = \pm$, we have $r_\alpha (-\vec{q}) v_\alpha^* (-\vec{q}) = R^*(-\vec{q}) v_\alpha^*(-\vec{q}) = R(\vec{q}) v_\alpha^*(-\vec{q})$. This shows that the spectrum of $R(\vec{q})$ -- and consequently of the polarization matrix $\Pi(\vec{q})$ -- is symmetric under $\vec{q} \to -\vec{q}$, such that the maxima of the function $p_{+}(\vec{q})$ must come in opposite pairs $\pm \vec{Q}$.

On the other hand, the invariance of $S$ under $U_g T_{\vec{a}_2/2}$ implies that
\begin{align}
	S_b^{(2)} 
	&\overset{!}{=} \sum_q b^\dagger (\vec{q} - \vec{b}_1/2, \ii\nu_n) \sigma_x R(\vec{q}, \ii\nu_n ) \sigma_x b (\vec{q} - \vec{b}_1/2, \ii\nu_n)
	\notag \\
	&= \sum_q b^\dagger (\vec{q}, \ii\nu_n) \sigma_x R(\vec{q} + \vec{b}_1/2, \ii\nu_n ) \sigma_x b (\vec{q}, \ii\nu_n),
\end{align}
where the real, off-diagonal Pauli matrix $\sigma_x$ implements the exchange of sublattices in Eq.~\eqref{eq:UgT}. We conclude that $R(\vec{q}, i\nu_n)$ and $R(\vec{q} + \vec{b}_1/2, \ii\nu_n )$ are related by a unitary transformation $R(\vec{q}, i\nu_n) = \sigma_x R(\vec{q} + \vec{b}_1/2, \ii\nu_n ) \sigma_x$ and thus have the same eigenvalues. As a result, if $S_b^{(2)}$ has a soft mode at momentum $Q_1 = (\vec{Q}_1, 0)$, there must be a degenerate one at $Q_2 = (\vec{Q}_1 + \vec{b}_1/2, 0)$.

When combined, the above symmetries explain why the function $p_{+} (\vec{q}, 0)$ has four degenerate global maxima at momenta $\pm \vec{Q}_1$ and $\pm \vec{Q}_2$ (see Fig.~4 of the main text). Since these are the only momenta that soften at $J_K = J_K^c$, the system admits a Landau free energy $\mathcal{F}_b$ that depends on four complex amplitudes $B_{\alpha s} = \expval{ b_{+} (s Q_\alpha) }$, with $\alpha = 1,2$ and $s = \pm$. Taking $\mathcal{T}$ and $U_g T_{\vec{a}_2/2}$ into account, one finds that
\begin{align}
	\mathcal{F}_b [B_{\alpha s}] &= \sum_{\alpha,s} \left( r \abs{ B_{\alpha s} }^2 + u \abs{ B_{\alpha s} }^4 \right)
	+ v \sum_{\alpha} \abs{B_{\alpha +}}^2 \abs{B_{\alpha -}}^2
	+ \sum_{s} \left( v' \abs{B_{1 s}}^2 \abs{B_{2 s}}^2 + v'' \abs{B_{1 s}}^2 \abs{B_{2 \bar{s}}}^2 \right)
	+ \cdots
	\label{eq:landau}
\end{align}
Apart from higher-order contributions in $B_{\alpha s}$, the ellipses in Eq.~\eqref{eq:landau} also represent quartic phase-sensitive terms
\begin{equation}
	\mathcal{F}_{b, \mathrm{phase}}^{(4)}
	=
	g_0 B_{1+}B_{1-}B_{2+}^*B_{2-}^*
	+
	g_1\sum_s B_{1s}^2(B_{2s}^*)^2
	+
	g_2 B_{1+}B_{2-}B_{1-}^*B_{2+}^*
	+\mathrm{c.c.},
	\label{eq:F4phase}
\end{equation}
including momentum-conserving ($g_0$) and Umklapp ($g_1$ and $g_2$) contributions.

In the interest of obtaining an overall qualitative picture of the Kondo-screened states that can emerge from the instability, we neglect Eq.~\eqref{eq:F4phase} and discuss some of the possible global minima of Eq.~\eqref{eq:landau}. We nonetheless emphasize that the stability conditions derived from this analysis can be affected by Eq.~\eqref{eq:F4phase}, which furthermore locks the relative complex phases of the different fields that condense. Table~\ref{tab:Fbminima} summarizes three distinct extrema of Eq.~\eqref{eq:landau}, corresponding to the condensation of (i) a single soft mode, (ii) a pair of soft modes with opposite wavevectors, and (iii) all four soft modes. In the remainder of this section, we discuss the implications of each of these scenarios.

\begin{table*}[t]
	\caption{Three types of solutions that can arise from the minimization of the Landau free energy $\mathcal{F}_b$ when $r<0$. The first four columns indicate which fields condense ($\star$) and which do not ($0$). The fifth column lists the values of the free energy at the corresponding stationary points, and the last column states the conditions under which each solution is a global minimum of $\mathcal{F}_b$.}
	\centering
	\begin{ruledtabular}
		\begin{tabular}{cccccc}
			$B_{1+}$ & $B_{1-}$ & $B_{2+}$ & $B_{2-}$ & $\mathcal{F}_b$ & Condition
			\\ \hline
			\rule[-2ex]{0pt}{6.5ex}
			$\star$ & $0$ & $0$ & $0$ & $- \dfrac{r^2}{4u}$ & $0 < 2u < v, v', v''$
			\\
			\rule[-2ex]{0pt}{6.5ex}
			$\star$ & $\star$ & $0$ & $0$ & $- \dfrac{r^2}{2u + v}$ &
			$-2u < v < \min\!\left\{2u,\;v',\;v'',\;v'+v''-2u\right\}$
			\\
			\rule[-2ex]{0pt}{6.5ex}
			$\star$ & $\star$ & $\star$ & $\star$ & $- \dfrac{2r^2}{2u + v + v' + v''}$ &
			$
			\max\!\left\{
			-(v+v'+v''),\;
			v'+v''-v,\;
			v+v''-v',\;
			v+v'-v''
			\right\} < 2u
			$
		\end{tabular}
	\end{ruledtabular}
	\label{tab:Fbminima}
\end{table*}

\subsubsection{Single-mode condensation: Intra-unit-cell modulations only}

Suppose that the coefficients $u,v,v',v''$ are such that $b_{+}$ only condenses at $\vec{Q}_1$, which is chosen spontaneously among the four degenerate solutions $\{ \pm \vec{Q}_1, \pm \vec{Q}_2 \}$. Then, the real-space configuration of the hybridization fields becomes
\begin{align}
		\expval{b_{ia}} &= B_{1+} w_a (\vec{Q}_1) e^{ \ii \vec{Q}_1 \cdot \vec{r}_{ia} },
		\label{eq:singleQcondensation}
\end{align}
where $w_a$ are sublattice weights that can be read off from Eq.~(12) in the main text. Importantly, the factor $e^{ \ii \vec{Q}_1 \cdot \vec{r}_{ia} }$ in Eq.~\eqref{eq:singleQcondensation} is a gauge-\emph{variant} quantity, and hence does \emph{not} imply a modulation of Kondo screening between different unit cells. Indeed, one can apply a local gauge transformation to the $f$ fermions that transfers the complex phase $e^{ \ii \vec{Q}_1 \cdot \vec{r}_{ia} }$ from the hybridization fields to the $f$ hopping amplitudes. This does not change the $f$ flux configuration on triangular plaquettes. However, if $\abs{w_A} \ne \abs{w_B}$, the state will also have asymmetric Kondo screening \emph{within} the two-site unit cell. The results in Fig.~3(b,c) show that this is generally the case.
 
\subsubsection{Condensation at a pair of opposite momenta: Single-$\vec{q}$ modulations}

On the other hand, if the quartic coefficients favor the simultaneous condensation of $b_{+}$ at a pair of opposite wavevectors $\pm \vec{Q}$, with $\vec{Q} = \vec{Q}_1$ or $\vec{Q}_2$, then
\begin{align}
	\expval{b_{ia}} &= B_{+} w_a (\vec{Q}) e^{ \ii \vec{Q} \cdot \vec{r}_{ia} } + B_{-} w_a (-\vec{Q}) e^{ -\ii \vec{Q} \cdot \vec{r}_{ia} }
	\notag \\
	&= B_{+} w_a (\vec{Q}) e^{ \ii \vec{Q} \cdot \vec{r}_{ia} } + B_{-} w_a^* (\vec{Q}) e^{ -\ii \vec{Q} \cdot \vec{r}_{ia} }.
	\label{eq:pmQcondensation}
\end{align}
Writing $w_a (\vec{Q}) = \abs{w_a (\vec{Q})} e^{\ii \varphi_{a\vec{Q}}}$ and $B_{\pm} = \abs{B_\pm} e^{\ii \varphi_\pm}$, this reduces to
\begin{align}
	\abs{ \expval{b_{ia}} }^2 &= \abs{ w_a (\vec{Q}) }^2 \left\{
	 \abs{B_{+}}^2 + \abs{B_{-}}^2 
	+ 2 \abs{B_+} \abs{B_-} \cos \left[
	2\left( \vec{Q} \cdot \vec{r}_{ia} + \varphi_{a\vec{Q}} \right) + \varphi_+ - \varphi_-
	\right]
	\right\},
\end{align}
which shows spatial modulations not only within the unit cell, but also between different unit cells with a wavevector $\vec{q}_\mathrm{HW} = 2\vec{Q}$. As in the previous case, the global complex phases of $\expval{b_{ia}}$ can be transferred to $f$ spinon hopping amplitudes via a gauge transformation. 

We note that it is in principle possible for the soft wavevectors to occur at $\vec{Q}_1 = \vec{0}$ and $\vec{Q}_2 = \vec{b}_1/2$, such that $\vec{q}_\mathrm{HW}$ becomes a reciprocal lattice vector. In this case, the system does not exhibit inter-unit-cell modulations and can only develop inhomogeneous Kondo screening via different sublattice weights $\abs{w_A} \ne \abs{w_B}$. Our considerations in the End Matter nonetheless indicate that this is a fine-tuned scenario which requires the bare $c$-electron Fermi surface to intersect the nodes in the dispersion of the Dirac spinons. Indeed, we only observed such an outcome in our explicit calculations for $n_c \approx 0.45$ (see Fig.~\ref{fig:cfs}).

\subsubsection{Condensation at all soft momenta: Multi-$\vec{q}$ modulations}

Another interesting scenario encompassed by the Landau free energy occurs when $b_{+}$ condenses at \emph{all four} momenta $\pm \vec{Q}_1$ and $\pm \vec{Q}_2$ at once. Taking the absolute value of
\begin{align}
	\expval{b_{ia}} &= \sum_{\alpha} \left[
	B_{\alpha+} w_a (\vec{Q}_\alpha) e^{ \ii \vec{Q}_\alpha \cdot \vec{r}_{ia} } + B_{\alpha-} w_a^* (\vec{Q_\alpha}) e^{ -\ii \vec{Q}_\alpha \cdot \vec{r}_{ia} }
	\right]
	\label{eq:fourQcondensation}
\end{align}
results in an intricate multi-$\vec{q}$ HW whose periodicity is governed by $\vec{Q}_1$ and $\vec{Q}_2$.


\bibliography{kondotr}